%% file: ms.tex
\documentclass[twocolumn, switch]{article} 

\usepackage{preprint}

\usepackage{enumitem}
\usepackage{amsmath, amsthm, amssymb, amsfonts}
\usepackage{bm}
\usepackage{amsmath}
\usepackage{graphicx}
\usepackage{subfigure}
\usepackage[algoruled,algo2e]{algorithm2e}

\usepackage{setspace}
\usepackage{amssymb}
\usepackage{booktabs}
\usepackage{array}
\usepackage{color}
\usepackage{multirow}
\usepackage{multicol}
\usepackage{threeparttable}
\usepackage{url}
\usepackage[page]{appendix}

\usepackage[numbers,square]{natbib}
\usepackage[utf8]{inputenc}	
\usepackage[T1]{fontenc}	
\usepackage{xcolor}		
\usepackage[colorlinks = true,
            linkcolor = blue,
            urlcolor  = blue,
            citecolor = blue,
            anchorcolor = black]{hyperref}	
\usepackage{booktabs} 		
\usepackage{nicefrac}		
\usepackage{microtype}		
\usepackage{lineno}		
\usepackage{float}			

\usepackage{newfloat}
\DeclareFloatingEnvironment[name={Supplementary Figure}]{suppfigure}
\usepackage{sidecap}
\sidecaptionvpos{figure}{c}

\usepackage{titlesec}
\titlespacing\section{0pt}{12pt plus 3pt minus 3pt}{1pt plus 1pt minus 1pt}
\titlespacing\subsection{0pt}{10pt plus 3pt minus 3pt}{1pt plus 1pt minus 1pt}
\titlespacing\subsubsection{0pt}{8pt plus 3pt minus 3pt}{1pt plus 1pt minus 1pt}

\usepackage{graphics}
\usepackage{hyperref}
\usepackage{color}
\usepackage{multirow}
\usepackage{hhline}
\usepackage{subfigure}
\usepackage{lipsum}

\newcommand{\etc}{\textit{etc}}

\newcommand{\etalq}{\textit{et al}. }

\newcommand{\wrtq}{\textit{w.r.t.} }
\newcommand{\cfq}{\textit{cf.} }

\newcommand{\ie}{\textit{i}.\textit{e}.}
\newcommand{\eg}{\textit{e}.\textit{g}.}

\title{Hierarchical Transformer with Spatio-Temporal Context Aggregation for Next Point-of-Interest Recommendation}

\usepackage{authblk}

\author[1]{Jiayi Xie}
\author[1,*]{Zhenzhong Chen}
\affil[1]{School of Remote Sensing and Information Engineering, Wuhan University}

\begin{document}

\twocolumn[ 
  \begin{@twocolumnfalse} 
  
\maketitle

\input{sections/0_abstract}
\vspace{0.4cm}

  \end{@twocolumnfalse} 
] 

\newcommand\blfootnote[1]{%
\begingroup
\renewcommand\thefootnote{}\footnote{#1}%
\addtocounter{footnote}{-1}%
\endgroup
}

{\blfootnote{* Corresponding author.}}
\input{sections/1_introduction}
\input{sections/2_relatedwork}
\input{sections/3_method}
\input{sections/4_experiments}
\input{sections/5_conclusion}

\bibliography{bib}

\end{document}

%% file: sections/0_abstract.tex
\begin{abstract}
Next point-of-interest (POI) recommendation is a critical task in location-based social networks, yet remains challenging due to a high degree of variation and personalization exhibited in user movements. In this work, we explore the latent hierarchical structure composed of multi-granularity short-term structural patterns in user check-in sequences. We propose a Spatio-Temporal context AggRegated Hierarchical Transformer (STAR-HiT) for next POI recommendation, which employs stacked hierarchical encoders to recursively encode the spatio-temporal context and explicitly locate subsequences of different granularities. More specifically, in each encoder, the global attention layer captures the spatio-temporal context of the sequence, while the local attention layer performed within each subsequence enhances subsequence modeling using the local context. The sequence partition layer infers positions and lengths of subsequences from the global context adaptively, such that semantics in subsequences can be well preserved. Finally, the subsequence aggregation layer fuses representations within each subsequence to form the corresponding subsequence representation, thereby generating a new sequence of higher-level granularity. The stacking of encoders captures the latent hierarchical structure of the check-in sequence, which is used to predict the next visiting POI. Extensive experiments on three public datasets demonstrate that the proposed model achieves superior performance whilst providing explanations for recommendations. Codes are available at \url{https://github.com/JennyXieJiayi/STAR-HiT}.
\end{abstract}

%% file: sections/1_introduction.tex
\section{Introduction}

With the prevalence of location-based services provided by applications such as Foursquare, Uber, Facebook, users are getting used to sharing their location-based experiences and acquiring location-aware services online. One of the most common location-based services is Point-of-Interest (POI) recommendation, which aims to predict the POI that is most likely to be visited by users. Next POI recommendation is a sub-field of POI recommendation that focuses on exploiting the user's historical trajectory to discover the potential sequential behavior patterns. On the one hand, next POI recommendation satisfies the personalized needs of users and alleviates information overload. On the other hand, it helps location-based service providers to provide intelligent location services, such as location-aware advertising, real-time Q\&A revolving around POIs, \etc. Therefore, next POI recommendation plays an increasingly important role in location-based services, and has attracted lots of attention from researchers in both academia and industry.

Next POI recommendation has been extensively studied \cite{tois22survey, neucom22survey}. Early studies focus on feature engineering (\eg, geospatial, temporal, social, content feature) and conventional machine learning models, such as Markov Chain (MC) based stochastic models and Matrix Factorization (MF) models to capture POI-POI transitions \cite{ijcai13successive, ijcai15prme}. These models rely on a strong assumption that the next POI for users to check-in is only determined by the last one or several check-ins. However, the next POI visited by a user is also highly correlated to other previous check-ins. As deep learning methods have shown promising performance compared to conventional methods, recent work turned to utilizing deep learning approaches to boost the performance of next POI recommendation \cite{nips13word2vec, aaai16strnn}. Among them, many studies adopt Recurrent Neural Network (RNN) to mine more complicated sequential patterns in long-term semantics \cite{aaai16strnn, aaai19stgn, ijcai20asppa}. They extend RNNs with the ability to effectively incorporate various contextual information, especially the spatio-temporal context. Some work also takes advantage of other deep learning techniques, such as attention mechanisms \cite{kdd20geosan, www21stan}, memory networks \cite{kdd19memory}, pre-training models \cite{aaai21pretraining}, meta-learning paradigms \cite{kdd21metalearning}, \etc. 

\begin{figure*}[htb]
    \centering
    \begin{minipage}[t]{0.5\linewidth}
		\centering
		\subfigure[Map]{
		\includegraphics[width=0.905\linewidth]{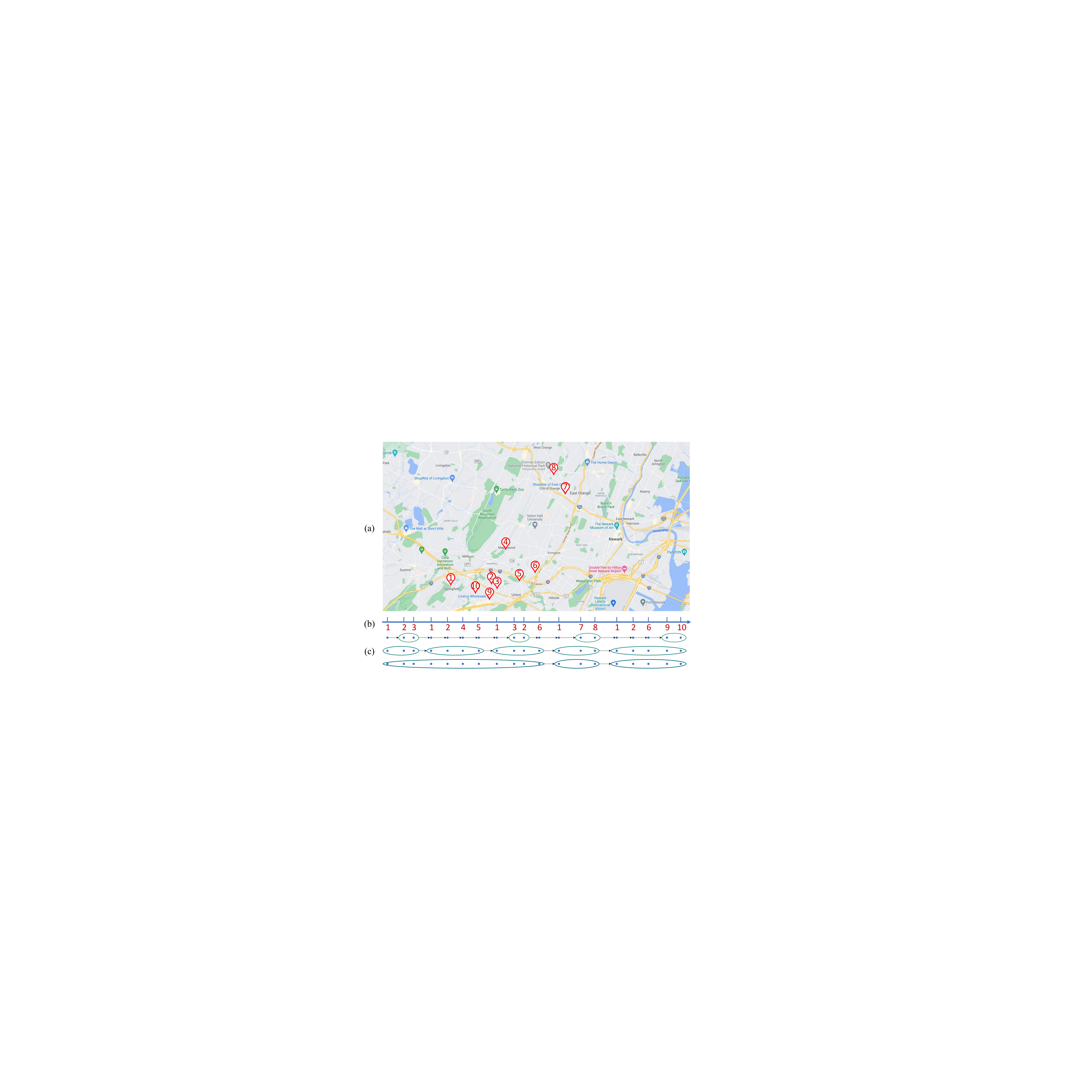}}
	\end{minipage}
	\begin{minipage}[t]{0.481\linewidth}
		\centering
		\vspace{22pt}
		\subfigure[Check-in Sequence]{
		\includegraphics[width=\linewidth]{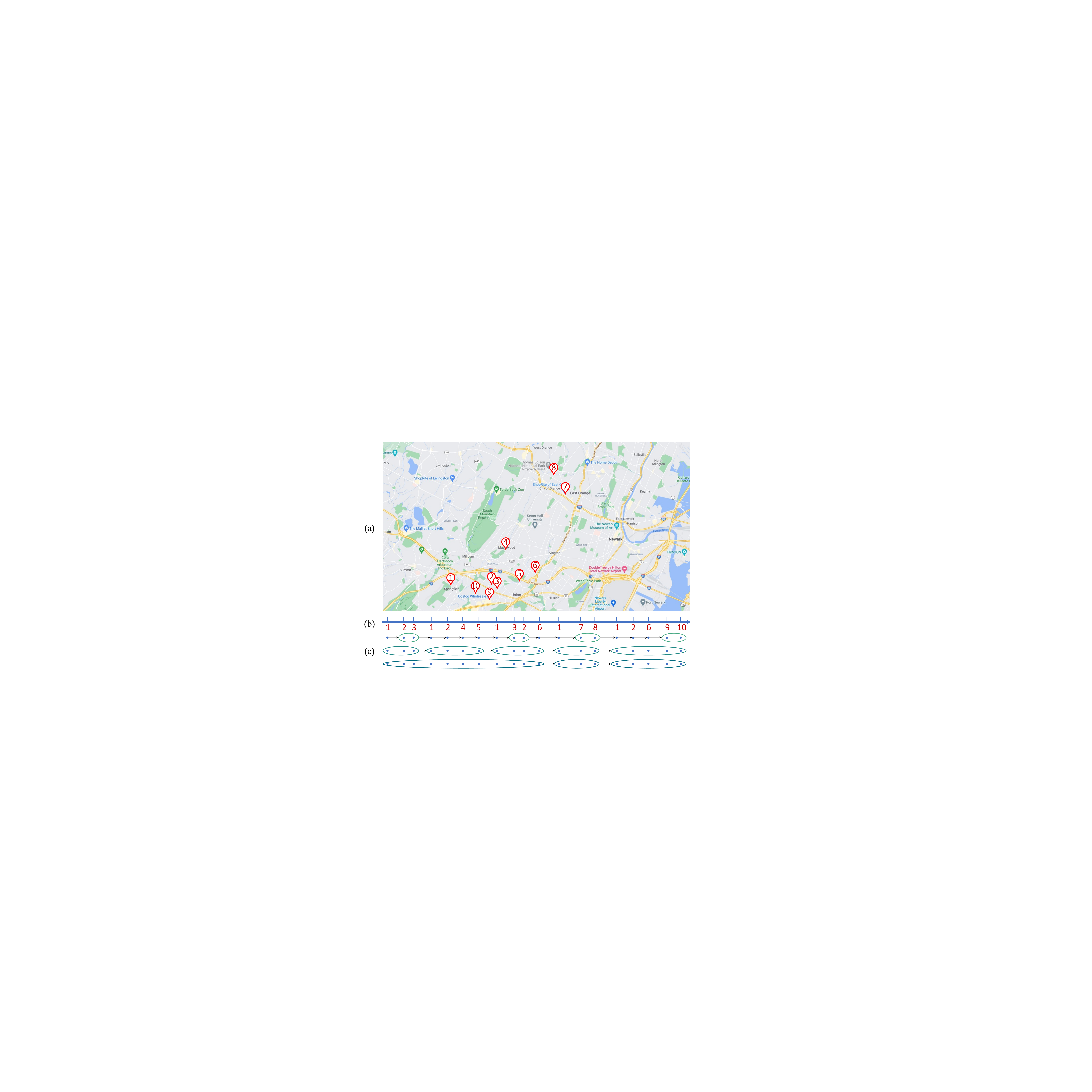}
		\label{fig:test}
		}\vspace{22pt}
		\subfigure[Subsequence Aggregation]{
		\includegraphics[width=\linewidth]{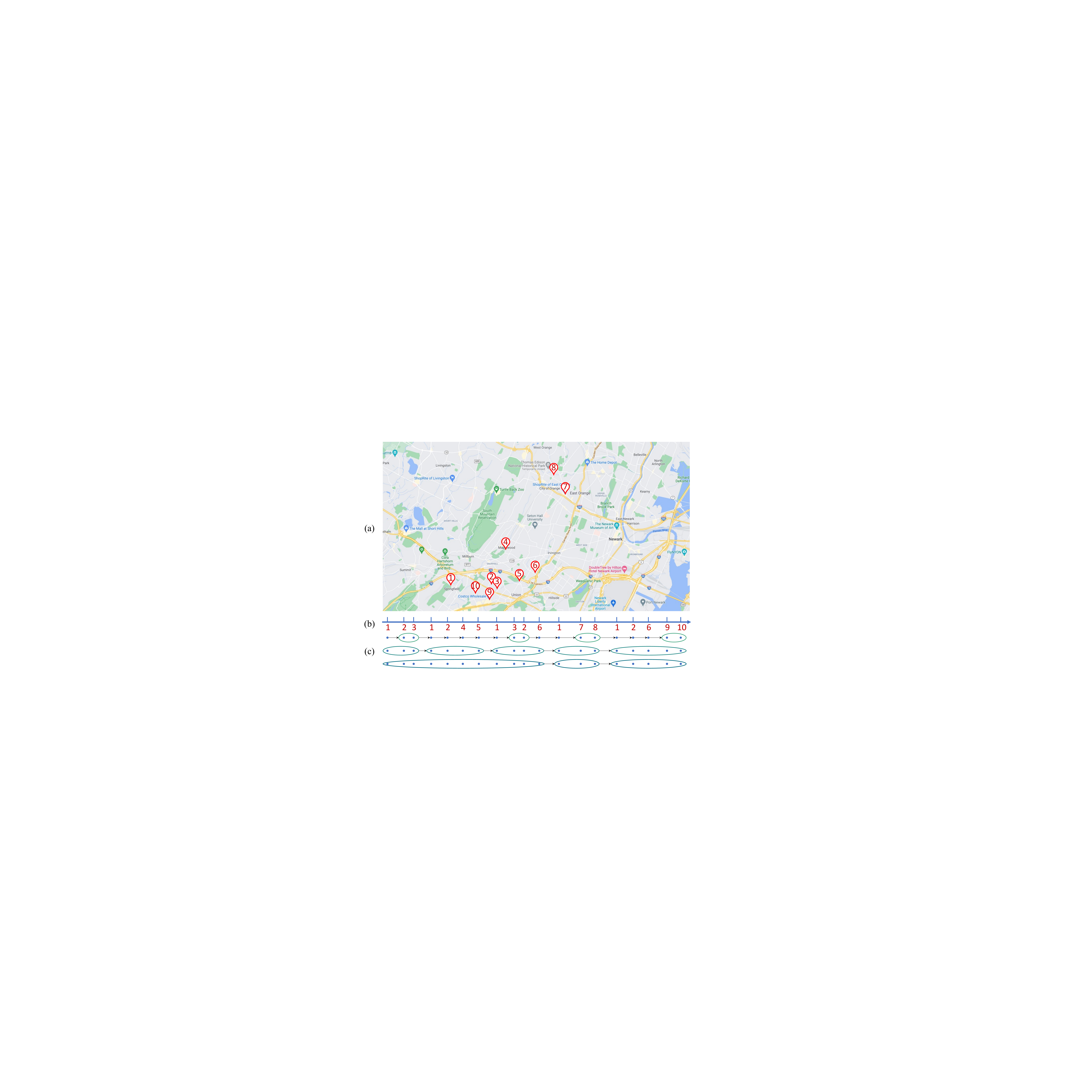}}
	\end{minipage}
\caption{A check-in sequence example.}
\label{fig:example}
\end{figure*}

Most previous methods assume that users only have preferences for some specific POIs, ignoring short-term structural patterns exhibited in user movements \cite{kdd11mf}. Such short-term structural patterns are highly personalized that can be caused by temporal regularities and geospatial constraints, resulting in multiple semantic subsequences at multi-level that comprise consecutive check-ins. The next visiting POI of the user could be correlated to several subsequences, which represent preferences beyond POI-level. As a special case shown in Figure \ref{fig:example}, a user often visits the area near her workplace (\ie, POI 2, 3) because of the convenience of visiting, instead of preferring a specific restaurant. Thus, other restaurants in that area could be the potential POI she would like to visit (\ie, POI 9). Such a high-level pattern can be inferred if she periodically visits the workplace area. Furthermore, the short-term sequential patterns could be multiple granularities. In particular, the user visits POIs in the home-workplace area in a regular manner (except POI 7, 8), such that the combination of check-in subsequences in the home area and workplace area could constitute the subsequence of a higher level granularity.

In order to exploit the aforementioned hierarchical structure of the check-in sequence, a naive way is to use a fixed-length subsequence embedding under an implicit assumption that the fixed-length subsequence is suitable for all short-term structural patterns in the check-in sequence. Such hard sequence partition neglects the personalized sequential behavior and could damage the semantic information. Nevertheless, it is challenging to identify and integrate multi-level semantic subsequences due to the difficulty of pre-defining granularities and lengths of subsequences \cite{ijcai20asppa}. How to comprehensively understand the overall sequential behavior patterns of users by discovering multi-level semantic subsequences remains to be explored. 

In light of the above, in this work, we aim to explore the latent hierarchical structure of the user movement by adaptively locating the semantic subsequence of multiple granularities in the check-in sequence. We propose a Spatio-Temporal context AggRegated Hierarchical Transformer (STAR-HiT) for next POI recommendation, which employs a stack of hierarchical encoders to jointly model the spatio-temporal context and capture the latent hierarchical structure in the check-in sequence. In particular, we design and stack the hierarchical encoder to recursively encode the spatio-temporal context and explicitly locate semantic subsequences, and generate subsequence representations to form a new sequence with a higher level of granularity. Note that every single check-in in the original check-in sequence can be regarded as the shortest subsequence. In each encoder, the global attention layer is utilized to capture the spatio-temporal correlations between subsequences, such that subsequences with similar sequential patterns could be associated. After the global context modeling, the sequence partition layer learns to adaptively locate next-level subsequences, followed by a local attention layer performed within each identified subsequence to enhance subsequence modeling using the corresponding local context. Finally, the subsequence aggregation layer fuses the representations in each next-level subsequence individually to form the subsequence representation, thereby generating a new sequence with a higher level of granularity. This sequence is then fed into the next encoder for further subsequence integrating and sequence abstraction. The intermediate representations in each encoder are learned to be expressive about the spatio-temporal context based on the global and local attention mechanisms; meanwhile, the sequence partition and subsequence aggregation operation can flexibly discover semantic subsequences of different positions and lengths. To summarize, this work makes the following main contributions:

\begin{enumerate}
    \item A novel next POI recommendation model STAR-HiT, consisting of stacked hierarchical encoders, is proposed to capture the latent hierarchical structure of the check-in sequence, from which the personalized movement pattern is revealed for the recommendation.
    \item A hierarchical encoder is designed to encode the spatio-temporal context and adaptively identify subsequences with different positions and lengths in the input sequence, then generate a sequence with a higher level of granularity. By stacking multiple hierarchical encoders, semantic subsequences of different granularities are recursively identified and integrated, so as to expose the overall hierarchical structure presented in the user movement.
    \item By capturing multi-level semantic subsequences that uncover the hierarchical structure, STAR-HiT guarantees the robustness and explainability of recommendations. Extensive experiments conducted on three public datasets demonstrate that our proposed STAR-HiT outperforms state-of-the-art models by a large margin whilst providing explanations for recommendations.
\end{enumerate}

The remaining of this paper is organized as follows: in Section 2, we review the related work. Section 3 expounds on the proposed STAR-HiT in detail, followed by experimental results with analysis on three public datasets in Section 4. Finally, Section 5 concludes the work.

%% file: sections/2_relatedwork.tex
\section{Related Work}

In this section, we first review the related work in the field of sequential recommendation and next POI recommendation. Then, we take a brief look at hierarchical Transformers applied to Natural Language Processing (NLP) and Computer Vision (CV).

\subsection{Sequential Recommendation}

Sequential recommendation mines behavior patterns in user action (\eg, click, watch, comment) sequences. Early work usually models an item-item transition pattern based on Markov chains. For example, Rendle \etalq \cite{fpmc} proposed the Factorized Personalized Markov Chains (FPMC) model that combines matrix factorization and Markov chains to incorporate both general preference and sequential behavior. Some studies follow this work and extend it to higher-order Markov chains \cite{fossil, recsys17mc}, where an $L$-order Markov chain is utilized to make predictions based on the $L$ previous actions. In general, Markov chain based models mainly focus on the latest short-term preference, performing relatively well in high-sparsity scenarios. With recent advances in deep learning, lots of work adopts neural network architectures, such as Convolutional Neural Networks (CNNs) \cite{wsdm18cnn, wsdm19cnn}, Recurrent Neural Networks (RNNs) \cite{iclr16gru, ijcai17timelstm}, Graph Neural Networks (GNNs) \cite{ijcai19gnn, aaai19gnn}, \etc. Among them, RNN is the most commonly used backbone, which encodes long-term dependencies in variable-length sequences. It performs well on dense datasets while exhibiting relatively poor performance on sparse datasets. Furthermore, some advanced deep learning techniques are utilized to enhance modeling capabilities, such as attention mechanisms \cite{ijcai18attn}, memory networks \cite{sigir18memory}, data augmentation \cite{sigir21augment}, denoising \cite{www22filter} and pre-training techniques \cite{www21pretraining}, \etc. More recently, Transformer-based approaches have shown remarkable performance in sequential recommendation. For instance, Kang \etalq \cite{icdm18sasrec} proposed a Self-Attention based Sequential Recommendation model (SASRec) that directly utilizes the stacked self-attention blocks in the vanilla Transformer to capture the correlation between every two items in the action sequence. SASRec outperforms state-of-the-art MC/CNN/RNN-based sequential recommendation methods on both sparse and dense datasets. Sun \etalq \cite{cikm19bert4rec} employed deep bidirectional self-attention to better model user behavior sequences. Later on, Li \etalq \cite{wsdn20tisasrec} improved SASRec by explicitly modeling the timestamps of interactions to emphasize the temporal influence. Wu \etalq \cite{recsys20ssept} introduced the personalization into SASRec by learning user embeddings with stochastic shared embeddings regularization. Moreover, Liu \etalq \cite{sigir21pretrainedtrm} alleviated the cold-start issue by augmenting short sequences with a pre-trained Transformer, which is trained on the reversed behavior sequences.

\subsection{Next POI Recommendation}

Next POI recommendation can be regarded as a special case of sequential recommendation, where geospatial influence is one of the most crucial context to be involved. Similar to the research on sequential recommendation, early studies on next POI recommendation mainly utilize Markov chains and matrix factorization \cite{ijcai13successive, ijcai15prme, aaai16infer}. Cheng \etalq \cite{ijcai13successive} extended FPMC model to FPMC-LR that models the personalized POI transition while considering users' movement constraint, namely, moving around a localized region. Feng \etalq \cite{ijcai15prme} further replaced the matrix factorization method with a metric embedding method and exploited the pair-wise ranking scheme to learn parameters. He \etalq \cite{aaai16infer} adopted a third-rank tensor to model the successive check-in behaviors and incorporated the softmax function to fuse the personalized Markov chain with users' latent behavior patterns.

With the surge of deep learning research, a large variety of next POI recommendation approaches leveraging different deep learning paradigms have been proposed. Among them, some work employs the word2vec framework \cite{nips13word2vec} to learn representations of POIs, which can reflect the contextual relationships of several continuously visited POIs \cite{aaai17poi2vec, www17geoteaser}. In addition, most existing models are based on RNNs, which have been widely applied to sequential data due to their powerful capability to exhibit temporal dynamics. For example, Liu \etalq \cite{aaai16strnn} adopted RNN to model the user's check-in sequence. The proposed ST-RNN model captures the spatial and temporal context with the time and distance transition matrices. Yang \etalq \cite{tois17neural} jointly modeled social networks and mobile trajectories by deriving user representations from social networks and adopting two different RNNs to encode long- and short-term sequential influence. Feng \etalq \cite{www18deepmove} enhanced GRU by utilizing attention mechanisms that capture the multi-level periodicity of users' mobility from long-range and sparse trajectories. Moreover, Guo \etalq \cite{aaai20arnn} employed LSTM as the recurrent layer, and designed a meta-path based random walk over a knowledge graph to discover location neighbors based on heterogeneous factors. Zhao \etalq \cite{aaai19stgn} extended the LSTM gating mechanism with the spatial and temporal gates to capture the user’s space and time preference. Zhao \etalq \cite{ijcai20asppa} introduced a binary boundary detector into RNN and modified the gating mechanism to learn the sequential patterns of semantic subsequences in the check-in sequence, then utilized a power-law attention mechanism to integrate the spatio-temporal context. Sun \etalq \cite{aaai20lstpm} combined a non-local network and a geo-dilated LSTM to leverage both long-term preference and geographical influence. Zang \etalq \cite{tois22cha} explored the category hierarchy of POIs to develop an attention-based knowledge graph for POI representation learning, and then proposed a spatial-temporal decay LSTM to capture the personalized behavior pattern. The above RNN-based models present many novel and enlightening contributions for introducing spatio-temporal context modeling into neural networks. Nevertheless, as aforementioned, RNN-based models need relatively dense data to train, while data sparsity is one of the most crucial problems in practical scenarios \cite{neucom22survey}.

Similar to sequential recommendation, some work also takes advantage of other deep learning techniques, such as attention mechanisms \cite{www21stan, kdd20geosan}, memory networks \cite{kdd19memory}, pre-training strategies \cite{aaai21pretraining}, meta-learning paradigms \cite{kdd21metalearning, tois22meta}, \etc. In particular, Zhou \etalq \cite{kdd19memory} proposed a Topic-Enhanced Memory Network that combines the topic model and memory network to jointly capture the global structure of latent patterns and local neighborhood-based features, while incorporating geographical influence by calculating a comprehensive geographical score. Kim \etalq \cite{kdd21metalearning} proposed an adaptive weighting scheme based on meta-learning to alleviate the class imbalance problem and noise of the input data. Cui \etalq \cite{tois22meta} jointly utilized a sequential knowledge graph and a meta-learning strategy to learn and optimize latent embeddings, thus modeling user check-in patterns while alleviating the data sparsity issue. As for attention-based models, Luo \etalq \cite{www21stan} employed a bi-layer attention architecture that allows global point-point interaction within the trajectory. The proposed method explicitly models the spatio-temporal context by embedding the spatio-temporal relation matrix of the trajectory. Lian \etalq \cite{kdd20geosan} developed a self-attention model to encode the check-in sequence, and encoded the hierarchical gridding of geospatial information with another self-attention based encoder. Lin \etalq \cite{aaai21pretraining} built a pre-training model to adaptively generate embeddings for locations based on their specific contextual neighbors. 

Although some studies have been devoted to investigating the short-term periodicity in the check-in sequence to optimize the network architecture \cite{www18deepmove, aaai20arnn, ijcai20asppa, tois22cha}, most previous work neglected the complicated yet structural patterns exhibited in user movements. In this work, we adopt the Transformer structure that has shown the convincing capability of dealing with long-term dependencies in sequences, and capture the latent hierarchical structure of user movements by adaptively discovering the multi-level semantic subsequences in an explicit way.

\subsection{Hierarchical Transformers}

Transformer architectures are based on the self-attention mechanism that learns the relationships among every element of a sequence, which can attend to complete sequences, thereby comprehensively understanding long-term context. Transformer was first proposed by Vaswani \etalq \cite{nips17trm} for machine translation, and has since become the state-of-the-art method in many NLP tasks. There are two mainstream approaches to enhancing the Transformer for better modeling longer-range dependencies with higher efficiency: variant self-attention mechanisms \cite{nips20bigbird, emnlp20} and hierarchical Transformer structures \cite{acl19hit, acl19hibert, cikm20hit, acl21hit}. Here we focus on the latter that leverages the natural hierarchical structure present in the syntax. Liu \etalq \cite{acl19hit} developed a multi-document summarization model and adopted a local Transformer layer and a global Transformer layer to encode the intra- and inter-paragraph contextual information, respectively. Yang \etalq \cite{cikm20hit} proposed a Siamese Multi-depth Transformer for document representation learning and matching, which contains sentence blocks and document context modeling. Wu \etalq \cite{acl21hit} effectively modeled long documents by a hierarchical Transformer following the sentence-document-sentence encoding strategy such that both sentence level and document level context could be integrated. 

The breakthroughs of Transformers in NLP have sparked great interest in CV tasks. Transformers for vision tasks usually segment the input image into a sequence of patches and capture long-range dependencies among patches. For example, Dosovitskiy \etalq \cite{vit} introduced the Vision Transformer for image classification, the first work directly applying the Transformer architecture and dispensing with convolutions entirely. To fit the Transformer, the input image is split into fix-size patches and linearly embedded into flat tokens to construct the input sequence. Liu \etalq \cite{swintrm} extended Vision Transformer by shifted windows, improving the efficiency by limiting the self-attention computation solely within each local window. They constructed hierarchical representations that start from small-sized patches and gradually merged neighboring patches in deeper layers. Wang \etalq \cite{pvt} improved Vision Transformer by incorporating the pyramid structure from convolutional neural networks. They utilized fine-to-coarse image patches to reduce the sequence length of Transformer as the network deepens, such that the input sequence elements can be set to pixel-level for dense prediction without increasing computational cost. Later on, Chen \etalq \cite{dpt} drew on the deformable convolution \cite{deformablecnn} and replaced the pre-defined patches with learnable patches in a data-driven way. As the proposed deformable patch embedding module splits the image into patches in a deformable way with learnable patch size and location, the semantics in patches can be well preserved.

Our work is partially inspired by the hierarchical enhancement of Transformer structures. In spite of the outstanding performance that these hierarchical Transformers have shown in various tasks, they cannot be directly used for next POI recommendation. Unlike the syntactic knowledge that helps pre-define the multi-scale subsequences for language modeling, the subsequences in the check-in sequence are personalized; thus, the extraction by a fixed length is unsuitable. Besides, the grid-topology spatial structure of visual data distinguishes them from check-in sequences, leading to the incompatibility of visual hierarchical Transformers to next POI recommendation. Nevertheless, these pioneering studies motivate us to extend Transformers to adaptively learn semantic multi-grained subsequences in the check-in sequence for better sequential behavior understanding.

%% file: sections/3_method.tex
\section{Methodology}

In this section, we first formulate the next POI recommendation task, then elaborate on the details of the proposed \textbf{S}patio-\textbf{T}emporal context \textbf{A}gg\textbf{R}egated \textbf{Hi}erarchical \textbf{T}ransformer (STAR-HiT). The notations mainly used in this article are listed in Table \ref{tab:notations}.

\begin{table*}
    \centering
    \caption{Notations Used in This Article}
    \begin{tabular}{l|p{10cm}}
    \toprule
    \makebox[0.15\linewidth][c]{{Variables}} &
    \makebox[0.7\linewidth][c]{{Description}} \\
    \midrule
    $m$, $n$, $L$ & the number of users, POIs, and the length of the check-in sequence \\
    $d$, $d_h$, $d_k$ & dimensions of latent representations \\
    $u$, $\mathcal{U}$ & a user and the user set \\
    $p$, $\mathcal{P}$ & a POI and the POI set \\
    $S_u$, $\mathcal{S}$ & the check-in sequence of the user $u$, the check-in sequence set\\
    $s^{(u)}_t$ & the $t$-th check-in of the user $u$ \\
    $g$, $\tau$ & geographic location and timestamp \\
    $\Delta^{\operatorname{S}} \in \mathbb{R}^{L\times L}$ & spatial relation matrix \\
    $\Delta^{\operatorname{T}}  \in \mathbb{R}^{L\times L}$ & temporal relation matrix \\
    $k$ & the initial length of the subsequence \\
    $l$ & the number of stacked hierarchical encoders \\
    $ \mathbf{E}(u) \in \mathbb{R}^{L\times d} $ & the representation matrix of the check-in sequence $S_u$ \\
    $\mathbf{E}(u)^{(l)} \in \mathbb{R}^{\lceil \frac{L}{k^{l}} \rceil \times d}$ & the representation matrix after $l$ hierarchical encoders of the check-in sequence $S_u$ \\

    \bottomrule
    \end{tabular}
    \label{tab:notations}
\end{table*}

\subsection{Problem Statement}

Let $ \mathcal{U} = \{u_1, u_2, \dots, u_m\} $, $ \mathcal{P} = \{p_1, p_2, \dots, p_n\} $ be the set of users and POIs, respectively. $\mathcal{S}=\{S_{1}, S_{2}, \ldots, S_{m}\}$ represents the set of user check-in sequences. For each user $u$, her check-in trajectory in chronological order is denoted as $ S_{u} = \{s^{(u)}_t \mid t = 1,2,\dots, L \} $, where $ s^{(u)}_t = (p_t, g_t, \tau_t) $ is the $t$-th check-in that user $u$ visits POI $ p_{t} \in \mathcal{P} $ with the geographic location $g_i = (latitude = \alpha_i, longitude = \beta_i)$ at timestamp $ \tau_t $. 

Next POI recommendation aims to recommend the POI that is most likely to be visited by the user at the next time step. Given a user $u$ with her check-in sequence $S_{u}$, the goal is to predict the next visiting POI $p_{t+1} \in \mathcal{P}$. The task can be formulated as estimating the personalized ranking score to the POI by:
\begin{equation}
    \hat{y}_{u,p} = f_{\Theta}(p \in \mathcal{P} \mid u, S_u),
\end{equation}

\noindent where $f_{\Theta}(\cdot)$ denotes the underlying model with parameters $\Theta$, and $\hat{y}_{u,p}$ is the predicted score for the check-in that user $u$ would like to visit POI $p$ at next time step. The top-k POIs ranked by predicted scores are the final recommendations.

\subsection{Overall Architecture}

\begin{figure*}
\centering
\includegraphics[width=0.99\linewidth]{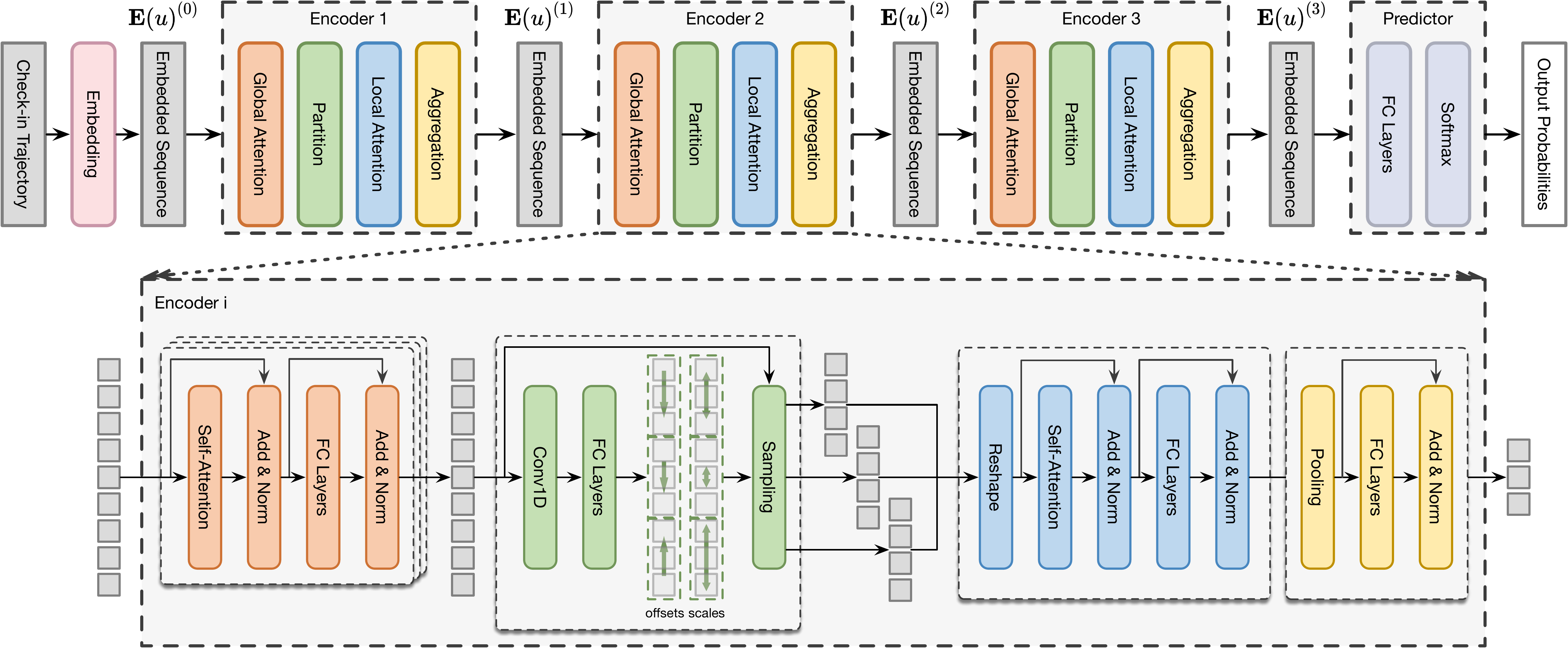}
\caption{The framework of the proposed STAR-HiT model.}
\label{fig:framework}
\end{figure*}

As shown in Figure \ref{fig:framework}, our proposed STAR-HiT consists of an embedding module, stacked hierarchical encoders, and the predictor. In particular, the embedding module embeds the POI and spatio-temporal context into latent representations to construct the spatio-temporal aware representation matrix of the check-in sequence. As for the stacked hierarchical encoders, the encoder adopts a hierarchical architecture that abstracts the input sequence to a compressive and expressive sequence. More specifically, the encoder first models the global spatio-temporal context within the entire sequence, or in other words, models the relationships between different subsequences learned in the previous encoder. Note that every single check-in in the very beginning check-in sequence is viewed as the shortest subsequence. Then the sequence is adaptively partitioned into next-level semantic subsequences with learnable positions and lengths, followed by the local context enhancement. Next-level subsequence representations are obtained by fusing their containing representations, which form the output sequence with a higher level of granularity. The hierarchical structure of the check-in sequence composed of multi-level semantic subsequences is learned based on stacked hierarchical encoders, with the purpose of comprehensively understanding users' overall sequential behavior patterns. Finally, a Multi-Layer Perceptron (MLP) based predictor is exploited to predict the check-in probabilities for users visiting POI. The details of STAR-HiT are elaborated as follows.

\subsection{Embedding Module}

The embedding module consists of two parts: a trajectory embedding layer and a spatio-temporal context embedding layer.

\subsubsection{Trajectory embedding layer}

The trajectory embedding layer encodes check-in POIs into latent representations of dimension $d$. We use $ e^{p_t} \in \mathbb{R}^d $ to denote the embedding of POI $p_t$ for user's $t$-th visiting. The embedding of each check-in sequence $S_{u}$ is represented as $\hat{\mathbf{E}}(u) = [e^{p_1}; e^{p_2}; \cdots; e^{p_L}] \in \mathbb{R}^{L\times d}$, where $L$ is the length of the sequence. The user embedding like \cite{kdd20geosan, www21stan} adopted is discarded, as the personalized information is already well preserved in the spatio-temporal context embedding to be introduced next.

\subsubsection{Spatio-temporal context embedding layer}

To leverage the spatio-temporal context of the check-in sequence, we first calculate the time intervals and geographical distances between every two visited POIs in a trajectory to construct the spatial-temporal relation matrices.

Following \cite{kdd20geosan}, for each check-in sequence $S_{u}$, we used $\Delta_{i, j}^{\operatorname{S}}=\operatorname{Haversine}(p_i, p_j)$ to obtain the distance between $i$-th and $j$-th visiting POI given their longitudes and latitudes\footnote{Haversine formula calculates the great-circle distance between two points on a sphere given their longitudes and latitudes.}. As for time intervals, instead of directly calculating the timestamp differences or uniformly grouping the values into discrete bins as \cite{aaai16strnn, www21stan} did, we resort to the scale of time intervals. Due to the extremely high variance of time intervals (\eg, from a few minutes to several years), we group the time interval into $M$ levels, namely, $\Delta_{i, j}^{\operatorname{T}} \in [1, M], M \in \mathbb{N}$, indicating the scale of the time interval between two check-ins ranging from within an hour to over a year. $M$ is set to $7$ in our implementation.

According to Tobler’s First Law of Geography \cite{tobler1970, sigir11tobler}, near POIs are more related than distant ones, which exhibits strong geographical influence. Inspired by \cite{sigir13tobler, ijcai20asppa}, we can consider that the  probability of visiting a pair of POIs $p_i$ and $p_j$ for user $u$ follows the power-law distribution as:
\begin{equation}
\operatorname{Pr}(p_i, p_j)=a \cdot \operatorname{D}(p_i, p_j)^{\lambda},
\label{eq:powerlaw}
\end{equation}

\noindent where $a$ and $k$ are learnable parameters of the power-law distribution, $D(p_i, p_j)$ is the distance between POI $p_i$ and $p_j$. We take the logarithm on both sides of Equation (\ref{eq:powerlaw}) as:
\begin{equation}
\operatorname{log}(\operatorname{Pr}(p_i, p_j)) = \operatorname{log}(a) + \lambda \cdot \operatorname{log} (\operatorname{D}(p_i, p_j)).
\label{eq:log_powerlaw}
\end{equation}

By leveraging the temporal influence, we extend Equation (\ref{eq:log_powerlaw}) by replacing the learnable $\operatorname{log}(a)$ with the temporal context as:
\begin{equation}
\operatorname{log}(\operatorname{Pr}(p_i, p_j)) = \operatorname{T}(p_i, p_j) + \lambda \cdot \operatorname{log}(\operatorname{D}(p_i, p_j)),
\label{eq:st_powerlaw}
\end{equation}

\noindent where $\operatorname{T}(p_i, p_j)$ represents the time interval between visiting $p_i$ and $p_j$. As such, both geographical and temporal influences are incorporated into the calculation of co-visiting probability. Furthermore, $k$ can be regarded as controlling the trade-off between geographical and temporal influence.

With the spatial relation matrix $\Delta^{\operatorname{S}}$ and temporal relation matrix $\Delta^{\operatorname{T}}$, we can obtain the spatio-temporal context embedding matrix $\mathbf{E}_c(u)$ of the check-in sequence $S_{u}$ as:
\begin{equation}
\mathbf{E}_c(u) = \Delta^{\operatorname{T}} + \boldsymbol{
\lambda} \cdot \operatorname{log}(\Delta^{\operatorname{S}}).
\end{equation}

We fuse all contextual information by concatenating the trajectory embedding and spatio-temporal context embedding, and then apply a linear transformation with learnable weights $\boldsymbol{W}^{E} \in \mathbb{R}^{(L+d) \times d}$ to obtain the final embedding of the check-in sequence:
\begin{equation}
\mathbf{E}(u) = \operatorname{Concat}([\hat{\mathbf{E}}(u); \mathbf{E}_c(u)]) \boldsymbol{W}^{E}.
\end{equation}

Since the self-attention mechanism contains no recurrence and no convolution to capture relative positions in the sequence like RNN, we follow \cite{nips17trm} to add positional encodings $\mathbf{P}$ into $\mathbf{E}(u)$, \ie, $\mathbf{E}(u) = \mathbf{E}(u) + \mathbf{P}$.

\subsection{Hierarchical Encoder}
In order to capture semantic subsequences of multiple granularities to obtain the hierarchical structure of the check-in sequence, the hierarchical encoder is supposed to extract semantic subsequences in the input sequence and derive their representations to generate a shorter output sequence with higher-level representations of behavior patterns. To this end, the proposed encoder comprises: 1) \textbf{global attention layer} that models the global context, 2) \textbf{sequence partition layer} that learns to locate semantic subsequences with different positions and lengths, 3) \textbf{local attention layer} that enhances subsequence modeling using the local context, 4) \textbf{subsequence aggregation layer} that obtains subsequence representations to construct a new sequence with an increased abstraction level. The details of the hierarchical encoder are shown at the bottom of Figure \ref{fig:framework}.

\subsubsection{Global Attention}

The global attention layer aims to learn the global context in the input sequence. Here, we adopt the encoder layer in the vanilla Transformer that contains two sub-layers, \ie, the multi-head self-attention sub-layer and position-wise feed-forward sub-layer. To briefly revisit Transformer, each encoder layer adaptively aggregates the values according to the attention weights that measure the compatibility of query-key pairs, where the query, key, value are all vectors transformed from the input representation. Such point-point interaction within the sequence allows the layer to capture long-term dependencies. Moreover, multi-head self-attention enables the layer to jointly attend to information from different representation subspaces at different positions.

In our case, the multi-head self-attention operation takes the embedding $\mathbf{E}(u)$ as input, linearly projects it into $h$ subspaces through distinct matrices, and then applies $h$ attention functions in parallel to produce the output representations, which are concatenated and once again projected. The whole multi-head self-attention operation can be summarized as follows:
\begin{equation}
    \operatorname{MSA}(\mathbf{E}(u)) = \operatorname{Concat}([\operatorname{SA}_1, \operatorname{SA}_2, \cdots, \operatorname{SA}_h])\boldsymbol{W}^{O},
\end{equation}

\noindent where
\begin{equation}
    \operatorname{SA}_i = \operatorname{Attention}(\mathbf{E}(u)\boldsymbol{W}^{Q}_i, \mathbf{E}(u)\boldsymbol{W}^{K}_i, \mathbf{E}(u)\boldsymbol{W}^{V}_i).
\end{equation}

\noindent The projection matrices for each head $\boldsymbol{W}^{Q}_i \in \mathbb{R}^{d \times d_h}, \boldsymbol{W}^{K}_i \in \mathbb{R}^{d \times d_h}, \boldsymbol{W}^{V}_i \in \mathbb{R}^{d \times d_h}$, $\boldsymbol{W}^{O} \in \mathbb{R}^{d \times d}$ are learnable parameters, where $d_h=d/h$. Note that the superscript $(l)$ indicating the $l$-th encoder is omitted above for simplicity. Here, the attention function is scaled dot-product attention, which is calculated as follows:
\begin{equation}
\operatorname{Attention}(\boldsymbol{Q}, \boldsymbol{K}, \boldsymbol{V})=\operatorname{softmax}(\frac{\boldsymbol{QK}^{T}}{\sqrt{d_h}}) \boldsymbol{V}.
\label{eq:attn}
\end{equation}

The multi-head self-attention sub-layer is followed by a position-wise feed-forward sub-layer, a fully connected two-layer feed-forward network applied to each position separately and identically. This consists of two linear transformations with a ReLU activation in between:
\begin{equation}
\operatorname{FFN}(x)=\operatorname{ReLU} (0, x \boldsymbol{W}_{1}+\boldsymbol{b}_{1}) \boldsymbol{W}_{2}+\boldsymbol{b}_{2},
\label{eq:ffn}
\end{equation}

\noindent where $\boldsymbol{W}_{1} \in \mathbb{R}^{d \times d_k}, \boldsymbol{W}_{2} \in \mathbb{R}^{d_k \times d}$ and $\boldsymbol{b}_{1} \in \mathbb{R}^{d_k}, \boldsymbol{b}_{2} \in \mathbb{R}^{d}, s.t.\ d_k > d$ are learnable parameters and shared across all positions. In addition, we also adopt residual connection \cite{resnet}, layer normalization \cite{layernorm}, and dropout regularization \cite{dropout} to refine the network structure \cite{nips17trm, cikm19bert4rec}. More specifically, we employ the residual connection around each of the two sub-layers, followed by layer normalization. The dropout is applied to the output of each sub-layer, before it is normalized. In summary, the sequence representation matrix after encoding the global context via the global attention layer is formulated as follows:
\begin{equation}
\begin{aligned}
\hat{\mathbf{E}}_G(u)&=\operatorname{LayerNorm}(\mathbf{E}(u)+\operatorname{Dropout}(\operatorname{MSA}(\mathbf{E}(u)))), \\
\mathbf{E}_G(u)&=\operatorname{LayerNorm}(\hat{\mathbf{E}}_G(u)+\operatorname{Dropout}(\operatorname{FFN}(\hat{\mathbf{E}}_G(u)))).
\end{aligned}
\end{equation}

\subsubsection{Sequence Partition}

In order to extract semantic subsequences in the check-in sequence, we first divide the sequence uniformly into $\lceil \frac{L}{k} \rceil$ non-overlapping subsequences of length $k$. Let $x_i$ denote the center coordinate of the $i$-th subsequence in the check-in sequence $S_u$, such that the location range of the subsequence in the input sequence is initialized as $(x_i - k/2, x_i + k/2)$. Next, we set the center coordinate $x_i$ and the length $k_i$ of each subsequence into learnable parameters, which can be inferred from the spatio-temporal context. In particular, inspired by \cite{dpt} for semantic patch learning in vision tasks, we predict the offset ${dx}_i$ and length $k_i$ based on the sequence representation $\mathbf{E}_G(u)$ as follows:
\begin{equation}
\begin{aligned}
{dx}_i&=\operatorname{Tanh}(w^1 \cdot f(\mathbf{E}_G(u)_i)), \\
k_i&=\operatorname{ReLU}(\operatorname{Tanh}(w^2 \cdot f(\mathbf{E}_G(u)_i))),
\end{aligned}
\end{equation}

\noindent where hyper-parameters $w^1, w^2$ control the weights of the offset and length to update the subsequence location, and $\mathbf{E}_G(u)_i$ is the representation matrix slicing corresponding to the $i$-th subsequence. $f(\cdot)$ denotes the feature extractor that learns the offset and length from subsequence representations. We follow \cite{dpt} and implement the feature extractor as a 1D convolution and a linear transformation with a ReLU activation in between.

Accordingly, the $i$-th learned subsequence is located in $(x_i+{dx}_i-k_i/2, x_i+{dx}_i+k_i/2)$. In this way, we can fully exploit the context to discover semantic subsequences. However, the subsequence with variable length makes it impractical to encode the intra-subsequence context subsequently. As a result, we introduce the sampling strategy to derive fixed-length subsequences. Given the location range of a subsequence in the input sequence $(x_{\operatorname{left}}, x_{\operatorname{right}})$, we first linearly interpolate $r$ points as $(x_1, x_2, \cdots, x_r)$ inside the subsequence. As the interpolated coordinates could be fractional, we then use the nearest-neighbor sampling to take representations closest to the corresponding coordinates, denoting as $\{\hat{e}^{[x_j]} \mid j=1, 2, \cdots, r\}$. Here we omit the superscript $(i)$ indicating the index of the subsequence for clarity. Finally, we concatenate the sampling representations as:
\begin{equation}
    \mathbf{E}_P(u)_i = \operatorname{Concat}([\hat{e}^{[x_1]}; \hat{e}^{[x_2]}; \cdots; \hat{e}^{[x_r]}]),
\end{equation}

\noindent so as to obtain the representation matrix of the $i$-th subsequence $\mathbf{E}_P(u)_i \in \mathbb{R}^{r\times d}$. In our implementation, the number of samples $r$ for each subsequence is set to the same as the initial length $k$ of subsequences. 

\subsubsection{Local Attention}

As semantic subsequences are identified, a local attention layer is applied afterward for encoding local contextual information within each subsequence. We use the same attention function as in Equation (\ref{eq:attn}) with only one head, since the subsequences are much shorter. We stack the current subsequence representations as $\mathbf{E}_P(u) \in \mathbb{R}^{\lceil \frac{L}{k} \rceil \times r\times d}$, such that the attention function for all subsequences can be computed in parallel as follows:
\begin{equation}
\operatorname{SA_{L}}(\mathbf{E}_P(u)) =  \operatorname{Attention}(\mathbf{E}_P(u)\boldsymbol{W}^{Q}_L, \mathbf{E}_P(u)\boldsymbol{W}^{K}_L, \mathbf{E}_P(u)\boldsymbol{W}^{V}_L)\boldsymbol{W}^{O}_L,
\end{equation}

\noindent where 
\begin{equation}
\operatorname{Attention}(\boldsymbol{Q}, \boldsymbol{K}, \boldsymbol{V})=\operatorname{softmax}(\frac{\boldsymbol{QK}^{T}}{\sqrt{d}}) \boldsymbol{V}.
\end{equation}

\noindent The projection matrices $\boldsymbol{W}^{Q}_L \in \mathbb{R}^{d\times d}$, $\boldsymbol{W}^{K}_L \in \mathbb{R}^{d\times d}$, $ \boldsymbol{W}^{V}_L \in \mathbb{R}^{d\times d}$ and $\boldsymbol{W}^{O}_L \in \mathbb{R}^{d\times d}$ are learnable parameters, where $\boldsymbol{W}^{Q}$, $\boldsymbol{W}^{K}$, $\boldsymbol{W}^{V}$ are shared across all subsequence representation matrices. Except for the local attention function calculated within each subsequence, the rest of the local attention layer is the same as the global attention layer, with individual parameters. Ultimately, the whole local attention layer can be described as follows:
\begin{equation}
\begin{aligned}
\hat{\mathbf{E}}_L(u)&=\operatorname{LayerNorm}(\mathbf{E}_P(u)+\operatorname{Dropout}(\operatorname{SA_{L}}(\mathbf{E}_P(u)))), \\
\mathbf{E}_L(u)&=\operatorname{LayerNorm}(\hat{\mathbf{E}}_L(u)+\operatorname{Dropout}(\operatorname{FFN}(\hat{\mathbf{E}}_L(u)))).
\end{aligned}
\end{equation}

\subsubsection{Subsequence Aggregation}
After context modeling via attention mechanisms and sequence partitioning into semantic subsequences, we gather the representations within each subsequence to obtain the corresponding subsequence representations, which constitute the output sequence. 

Given the representation matrix $\mathbf{E}_L(u)$, the representation of the $i$-th subsequence is obtained by the average pooling of the representations of all representations it contains, which is formulated as:
\begin{equation}
    \hat{\mathbf{E}}_A(u)_i = \frac{1}{r} \sum_{j=ir}^{(i+1)r}\mathbf{E}_L(u)_{i,j},
\end{equation}

\noindent followed by a fully connected two-layer feed-forward network as in Equation (\ref{eq:ffn}), with the aforementioned techniques that ease the training. Accordingly, the output of the subsequence aggregation layer, as well as the output of the hierarchical encoder, is obtained by:
\begin{equation}
    \mathbf{E}_A(u) = \operatorname{LayerNorm}(\hat{\mathbf{E}}_A(u)+\operatorname{Dropout}(\operatorname{FFN}(\hat{\mathbf{E}}_A(u)))).
\end{equation}

By now, the structure of the proposed hierarchical encoder is fully specified. Through the hierarchical encoder, the global context of the sequence and local contexts of semantic subsequences are well involved, while the input sequence $\mathbf{E}(u)^{(l-1)} \in \mathbb{R}^{\lceil \frac{L}{k^{(l-1)}} \rceil \times d}$ of the $l$-th encoder are abstracted to the output sequence $\mathbf{E}(u)^{(l)} = \mathbf{E}_A(u)^{(l)} \in \mathbb{R}^{\lceil \frac{L}{k^{l}} \rceil \times d}$, where $(l)$ indicates the $l$-th encoder. Besides, semantics subsequences in each encoder are identified by the learned positions and lengths.

\subsubsection{Stacking Encoders}
In order to model the latent hierarchical structure of the sequential behavior pattern from the check-in sequence, we stack the hierarchical encoders to recursively partition the input sequence into multiple semantic subsequences and aggregate them to form the output sequence with different levels of granularity. 

As aforementioned, the length of sequence is reduced from $L$ to $ \lceil \frac{L}{k^l} \rceil$ after $l$ hierarchical encoders. The output of the hierarchical encoder stack would be of shorter length while highly informative about the personalized behavior sequential pattern; meanwhile, each encoder is learned to be capable of discovering semantic subsequences with different levels of granularity. The number of stacked encoders $l$ will be discussed in Section \ref{sec:params}.

\subsection{Prediction}
To predict the next check-in POI, we first obtain the user representation $\boldsymbol{U}_u$ of user $u$ by summing up the output $\mathbf{E}(u)^{(l)}$ after the stack of $l$ hierarchical encoders, which constitutes the user representation matrix $\boldsymbol{U} \in \mathbb{R}^{m\times d}$ that represents the personalized sequential behavior patterns of all users.

A commonly used prediction layer adopted the matching function as follows:
\begin{equation}
    \hat{y}_{u,p} = \boldsymbol{U}_u^\top\boldsymbol{P}_p,
\end{equation}

\noindent where $\boldsymbol{P}_p$ is the embedding of POI $p \in \mathcal{P}$ \cite{icdm18sasrec, www21stan}. However, further adding the POI embeddings to calculate the matching score degrades the performance of the proposed STAR-HiT to some extent. The main reason may lie in the difference between the personalized representation encoded by encoders and the general representation obtained by the embedding layer. Therefore, the matching function is not adopted in our implementation. Instead, we directly use a linear transformation with learnable weight $\boldsymbol{W}^P \in \mathbb{R}^{d \times n}$ and a softmax function to convert the output into predicted next POI probabilities as follows:
\begin{equation}
    \hat{\boldsymbol{y}}_{u} = \operatorname{softmax}(\boldsymbol{U}_u\boldsymbol{W}^P),
\end{equation}

\noindent where $\hat{\boldsymbol{y}}_{u} = [\hat{y}_{u,1}, \hat{y}_{u,2}, \cdots, \hat{y}_{u,n}]$ is the predicted score of user $u$ to visit the candidate POIs.

To train the model, we adopt the cross-entropy loss to optimize parameters as:
\begin{equation}
    \mathcal{L} = -\sum_{S_u \in \mathcal{S}_{\operatorname{training}}}(\log \hat{y}_{u,i}+\sum_{j \in \mathcal{P}, j \neq i} \log (1-\hat{y}_{u,j})),
\end{equation}

\noindent where $\mathcal{S}_{\operatorname{training}}$ is the training set of check-in sequences.

%% file: sections/4_experiments.tex
\section{Experiments} 

In this section, we conduct experiments to show the effectiveness of the proposed STAR-HiT. Specifically, we aim to answer the following research questions:

\begin{itemize}
    \item \textbf{RQ1: }How does STAR-HiT perform compared to state-of-the-art methods on next POI recommendation?
    \item \textbf{RQ2: }How do different designs of STAR-HiT influence the performance?
    \item \textbf{RQ3: }Can STAR-HiT capture the latent hierarchical structure present in check-in sequences?
\end{itemize}

In what follows, we first introduce datasets, evaluation metrics and compared methods, followed by answering the above questions. In particular, we present the performance comparison with analysis among STAR-HiT and state-of-the-art baseline methods. Then, we explore how the hyper-parameter settings influence STAR-HiT, in terms of the initial length of subsequences, the number of stacked hierarchical encoders, the dimension of representations, \etc. In addition, we examine the effect of different modules in STAR-HiT, in terms of four layers in the proposed hierarchical encoder and the learnable localization of subsequences. We also validate the capability of STAR-HiT to capture the personalized latent hierarchical structure of the check-in sequence by the case study.

\subsection{Experimental Settings}

\subsubsection{Datasets}
To evaluate the effectiveness of STAR-HiT, we conduct experiments on three publicly available datasets: Foursquare NYC, Foursquare US, and Gowalla. 

\begin{itemize}
    \item \textbf{Foursquare NYC:} Foursquare NYC \cite{tsmc15foursquarenyc} is a widely used dataset for POI recommendation, which contains check-ins in New York city collected from April 2012 to February 2013.
    \item \textbf{Foursquare US:} This dataset is a subset of a long-term global-scale check-in dataset collected from Foursquare \cite{tist16foursquareglobal}. Following \cite{ijcai20asppa}, we use check-in data within the United States (except Alaska and Hawaii), and rename the dataset as Foursquare US. 
    \item \textbf{Gowalla:} This is a check-in dataset obtained from Gowalla \cite{kdd11mf} over the period of February 2009 to October 2010.
\end{itemize}
    
Table \ref{tab:dataset_statistics} summarizes the statistics of three datasets, where \textit{Revisit Frequency} refers to the ratio of the total number of check-ins to the number of visited POIs of a user. As for \textit{Revisit Ratio}, it depicts the ratio of the number of repeated check-ins for the same POIs to the total number of check-ins of a user. Both of the above metrics measure re-visits in the dataset, which is complementary information to the sparsity. We report the average Revisit Frequency and Revisit Ratio of all users.

For each dataset, we filter out users and POIs with fewer than 10 check-ins as previous work \cite{ijcai20asppa} did. For the check-in sequence of each user, we set the maximum sequence length as $L_{\operatorname{max}}$. Then we slide the fixed-length window on the original check-in trajectory to obtain sequence slices when $L > L_{\operatorname{max}}$, otherwise padding with zeros to the right to construct the sequence of length $L_{\operatorname{max}}$. Since the number of users in Foursquare NYC is much smaller, $L_{\operatorname{max}}$ is set to 100 for Foursquare NYC and 128 for others. In order to simulate the real-world next POI recommendation scenario, we rank the check-in sequence of each user in chronological order and split the dataset into training (80\%), validation (10\%), and test (10\%) sets.

\begin{table*}[]
\centering
\caption{Statistics of Datasets}
\begin{tabular}{lccc}
\toprule
 & \makebox[0.2\linewidth][c]{{Foursquare NYC}} &
\makebox[0.2\linewidth][c]{{Foursquare US}} &
\makebox[0.2\linewidth][c]{{Gowalla}} \\
\midrule
\# users & 1,083 & 30,410 & 51,989 \\
\# POIs & 5,135 & 79,580 & 131,282 \\
\# check-ins & 147,938 & 2,440,233 & 3,365,444 \\ 
Avg. POIs per user & 137 & 80 & 65 \\
Avg. users per POI & 29 & 31 & 26 \\
Revisit Frequency & 4.89 & 2.88 & 2.56 \\
Revisit Ratio & 36.45\% & 26.72\% & 28.06\% \\
Sparsity & 97.34\% & 99.90\% & 99.95\% \\
\bottomrule
\end{tabular}
\label{tab:dataset_statistics}
\end{table*}
    
\subsubsection{Evaluation Metrics}

To evaluate the performance of next POI recommendation methods, we adopt two widely used evaluation protocols for recommendation systems \cite{www17ncf}: HR@K and NDCG@K. For each test sequence, we predict the probabilities of candidate POIs $p \in \mathcal{P}$ and recommend the top-$K$ POIs. Hit Ratio (HR) measures whether the true next visiting POI is present on the top-$K$ ranked list, while Normalized Discounted Cumulative Gain (NDCG) further emphasizes the position of the hit by assigning higher weights to hits at topper ranks. We set $K = 5$ and $K = 10$ and report the average metrics for all sequences in the test set.

\subsubsection{Compared Methods}

We compare our proposed STAR-HiT with a statistical model (MFLM), RNN-based models (GRU4Rec, Time-LSTM, STRNN, STGN), attention-based model (STAN), and Transformer-based models (SASRec, SSE-PT, TiSASRec, GeoSAN), as follows:

\begin{itemize}
    \item \textbf{MFLM:} Most Frequented Location Model \cite{kdd11mf} is a statistical-based model, which calculates the probability of a user visiting the POI based on the statistics of her previous check-ins. It captures the periodic check-in habit of the user.
    \item \textbf{GRU4Rec:} GRU4Rec \cite{iclr16gru} models the user action sequence for session-based recommendation utilizing Gated Recurrent Unit (GRU), a variant of RNN. We adopt a two-layer GRU for modeling check-in sequences.
    \item \textbf{Time-LSTM:} Time-LSTM \cite{ijcai17timelstm} is a variant of LSTM that improves the modeling of sequential patterns by explicitly incorporating time intervals with designed time gates. We adopt the third version proposed in the paper since it achieves the best performance in our experiments. 
    \item \textbf{STRNN:} Spatial Temporal Recurrent Neural Networks \cite{aaai16strnn} improves RNN for check-in sequence modeling by capturing local temporal and spatial contexts with time and distance transition matrices.
    \item \textbf{STGN:} Spatio-temporal Gated Network \cite{aaai19stgn} extends the gating mechanism of LSTM with four spatial-temporal gates to capture the user’s both long-term and short-term space and time preference.
    \item \textbf{STAN:} Spatio-Temporal Attention Network \cite{www21stan} uses a bi-layer attention architecture to explicitly exploit point-to-point spatio-temporal correlations in check-in sequences, so that correlations of non-adjacent locations and non-consecutive check-ins are well incorporated for understanding user behavior.
    \item \textbf{SASRec:} Self-Attention based Sequential Recommendation \cite{icdm18sasrec} directly implement the Transformer \cite{nips17trm} architecture, taking advantage of the self-attention mechanism to adaptively assign high weights to relatively few but relevant actions for recommendations. 
    \item \textbf{SSE-PT:} Personalized Transformer with Stochastic Shared Embeddings (SSE) regularization \cite{recsys20ssept} extends SASRec by introducing personalization into the model. The usage of SSE regularization prevents the model from overfitting after leveraging user embeddings.
    \item \textbf{TiSASRec:} Time interval aware Self-Attention based Sequential Recommendation \cite{wsdn20tisasrec} extends SASRec by modeling both absolute positions of items and personalized time intervals between them in sequences.
    \item \textbf{GeoSAN:} Geography-aware sequential recommender based on the Self-Attention Network \cite{kdd20geosan} explicitly utilizes the time of check-ins and GPS positions of POIs. In particular, a self-attention based geography encoder is designed to encode the geographical information of each POI in the sequence.
\end{itemize}

\subsubsection{Implementation Details}

We implement the compared methods following the original settings. It should be noted that the experimental results of STAN on Foursquare US and Gowalla are ignored, due to its extremely high memory usage for the distance matrix of all POIs in large-scale datasets. Besides, we remove the geography-aware negative sampler in GeoSAN for a fair comparison, as it is not performed by other methods. We employ the BPR loss \cite{bprloss} for optimizing the RNN-based models, since the cross-entropy loss generally leads to poor performance. As for our proposed STAR-HiT, the embedding dimension $d$ is set to 68 and the hidden dimension $d_k$ in the feed-forward networks is set to twice the embedding dimension, which is 128. The weights $w^1, w^2$ that control the offset and scale to update the subsequence location are both set to 1, and the dropout ratio is set to 0.2. Inspired by \cite{nips17trm}, we train the model using the Adam optimizer \cite{adam} with $\beta_{1}=0.9, \beta_{2}=0.98, \epsilon=10^{-9}$, and the learning rate $\operatorname{l\_rate}$ is varied over the course of training as: $\operatorname{l\_rate}=\lambda \cdot d^{-0.5} \cdot \min (\operatorname{num\_step}^{-0.5}, \operatorname{num\_step} \cdot \operatorname{warmup\_step}^{-1.5})$, where $d$ is the embedding dimension, $\operatorname{num\_step}$ is the number of training steps, the coefficient $\lambda$ controls the overall learning rate which is set to 1, $\operatorname{warmup\_step}$ refers to the training step with the peak learning rate and is set to 400. We use the Xavier initialization \cite{xavier} to initialize model parameters. For all models, we use the default number of training epochs of 200 for Foursquare NYC and 300 for others, and the default mini-batch size of 128. Our model is implemented in PyTorch and available at \url{https://github.com/JennyXieJiayi/STAR-HiT}.

\subsection{Performance Comparison (RQ1)}

\begin{table*}[htb]
    \centering
    \caption{Performance Comparison with Baseline Methods}
    \label{tab:perf_baselines}
    \resizebox{\textwidth}{!}{
    \begin{tabular}{lcccccccccccc}
    \toprule
    & \multicolumn{4}{c}{Foursquare NYC}  &  \multicolumn{4}{c}{Foursquare US}   &  \multicolumn{4}{c}{Gowalla}         \\
    \cmidrule(lr){2-5} \cmidrule(lr){6-9} \cmidrule(lr){10-13}
    & H@5 & H@10 & N@5 & N@10 
    & H@5 & H@10 & N@5 & N@10    
    & H@5 & H@10 & N@5 & N@10 \\ \midrule
    MFLM & 
    0.1412 & 0.1421 & 0.1316 & 0.1319 &
    0.1386 & 0.1394 & 0.1296 & 0.1299 & 
    0.1176 & 0.1176 & 0.1106 & 0.1106 \\
    GRU4Rec & 
    0.1421 & 0.2335 & 0.1006 & 0.1310 & 
    0.1412 & 0.1908 & 0.1074 & 0.1227 & 
    0.1459 & 0.1983 & 0.1057 & 0.1227 \\
    Time-LSTM & 
    0.1584 & 0.2534 & 0.1085 & 0.1429 & 
    0.1426 & 0.1928 & 0.1091 & 0.1247 & 
    0.1465 & 0.2002 & 0.1076 & 0.1243 \\
    STRNN & 
    0.1475 & 0.2425 & 0.1006 & 0.1317 & 
    0.1738 & 0.2424 & 0.1314 & 0.1533 & 
    0.1474 & 0.2045 & 0.1078 & 0.1250 \\
    STGN & 
    0.1548 & 0.2561 & 0.1132 & 0.1452 & 
    0.1803 & 0.2538 & 0.1341 & 0.1588 & 
    0.1467 & 0.2088 & 0.1070 & 0.1264 \\
    STAN & 
    0.3587 & 0.5112 & 0.2506 & 0.3008 & 
    - & - & - & - & 
    - & - & - & - \\
    SASRec & 
    0.3439 & 0.4597 & 0.1719 & 0.1882 & 
    0.2727 & 0.3765 & 0.1276 & 0.1356 & 
    0.2550 & 0.3635 & 0.1169 & 0.1253 \\
    SSE-PT & 
    0.3719 & 0.4950 & 0.2362 & 0.2751 & 
    0.3181 & 0.3829 & 0.2029 & 0.2056 & 
    0.3049 & 0.4050 & 0.1596 & 0.1667 \\
    TiSASRec & 
    0.3665 & 0.4860 & 0.2621 & 0.3008 & 
    0.2803 & 0.3834 & 0.1419 & 0.1491 & 
    0.2926 & 0.3851 & 0.1462 & 0.1522 \\
    GeoSAN & 
    \underline{0.4847} & \underline{0.5571} & \underline{0.3396} & \underline{0.3630} & 
    \underline{0.4100} & \underline{0.4991} & \underline{0.3263} & \underline{0.3551} & 
    \underline{0.3349} & \underline{0.4183} & \underline{0.2583} & \underline{0.2855} \\
    STAR-HiT & 
    \textbf{0.5991}&\textbf{0.6597}&\textbf{0.5186}&\textbf{0.5385} & 
    \textbf{0.6968}&\textbf{0.7296}&\textbf{0.6381}&\textbf{0.6486} & 
    \textbf{0.4497}&\textbf{0.4929}&\textbf{0.3921}&\textbf{0.4057} \\ 
    \midrule
    \textit{Improv.} & 
    23.60\% & 18.42\% & 52.69\% & 48.32\% & 
    69.94\% & 46.18\% & 95.58\% & 82.65\% & 
    34.29\% & 17.82\% & 51.79\% & 42.13\% \\ \bottomrule 
    \end{tabular}}
    \end{table*}

The performance comparison with baseline methods are illustrated in Table \ref{tab:perf_baselines}. We have the following observations:

\begin{itemize}
    \item The proposed STAR-HiT achieves the best performance among all the compared methods. In particular, STAR-HiT improves the performance over the strongest baseline, \ie, GeoSAN, in terms of HR@5 by 23.6\%, 69.94\%, 34.29\% in Foursquare NYC, Foursquare US, and Gowalla, respectively. Moreover, the corresponding performance improvements in terms of NDCG@5 are 52.69\%, 95.58\%, and 51.79\%, respectively. By stacked hierarchical encoders, STAR-HiT benefits from spatio-temporal context modeling and multi-granularity semantic subsequences discovering in an explicit manner, so as to model the inherent hierarchical structure exhibited in check-in sequences. This verifies the significance of modeling the hierarchical structure of check-in sequences to improve the recommendations.
    \item The statistical model MFLM achieves relatively poor performance on Foursquare US and Gowalla, while performing comparable to RNN-based models (\ie, GRU4Rec, Time-LSTM, STRNN, STGN) on Foursquare NYC. The reason may lie in that there are more revisits for users in Foursquare NYC than others (\cfq Table \ref{tab:dataset_statistics}). Combining the subtle differences of MFLM performing in terms of HRs and NDCGs, we can conclude that MFLM is suitable for users with periodic check-in patterns, while unable to deal with those with a certain degree of flexible check-in patterns.
    \item STAN and Transformer-based models (\ie, SASRec, SSE-PT, TiSASRec, GeoSAN, STAR-HiT) with the self-attention mechanism as the major component, consistently outperform RNN-based models. The recurrent neural networks handle the check-in sequence by recursively encoding previous check-ins into the internal memory as a whole, such that correlations between non-consecutive check-ins and long-term semantics are underestimated. Instead, the self-attention mechanism allows capturing the correlations of any two check-ins in an explicit way, despite whether they are consecutive.
    \item Among RNN-based methods, models that consider at least one of the spatial and temporal contexts (\ie, Time-LSTM, STRNN, STGN) perform better than others (\ie, GRU4Rec). Likewise, GeoSAN and STAR-HiT that embed the spatio-temporal context outperform SASRec, SSE-PT, and TiSASRec designed for traditional items. These results highlight the importance of spatio-temporal context modeling.
\end{itemize}

\subsection{Study of STAR-HiT (RQ2)}

To get deep insights into the design of STAR-HiT, we first explore how different settings influence the performance, in terms of the initial length of subsequences, the number of stacked hierarchical encoders, the embedding dimension. Then, we analyze the effectiveness of the various components by conducting an ablation study.

\subsubsection{Parameter Analysis}
\label{sec:params}

\begin{figure*}[htb]
    \centering
    \begin{minipage}[t]{0.32\linewidth}
        \centering
        \subfigure{
        \includegraphics[width=0.96\linewidth]{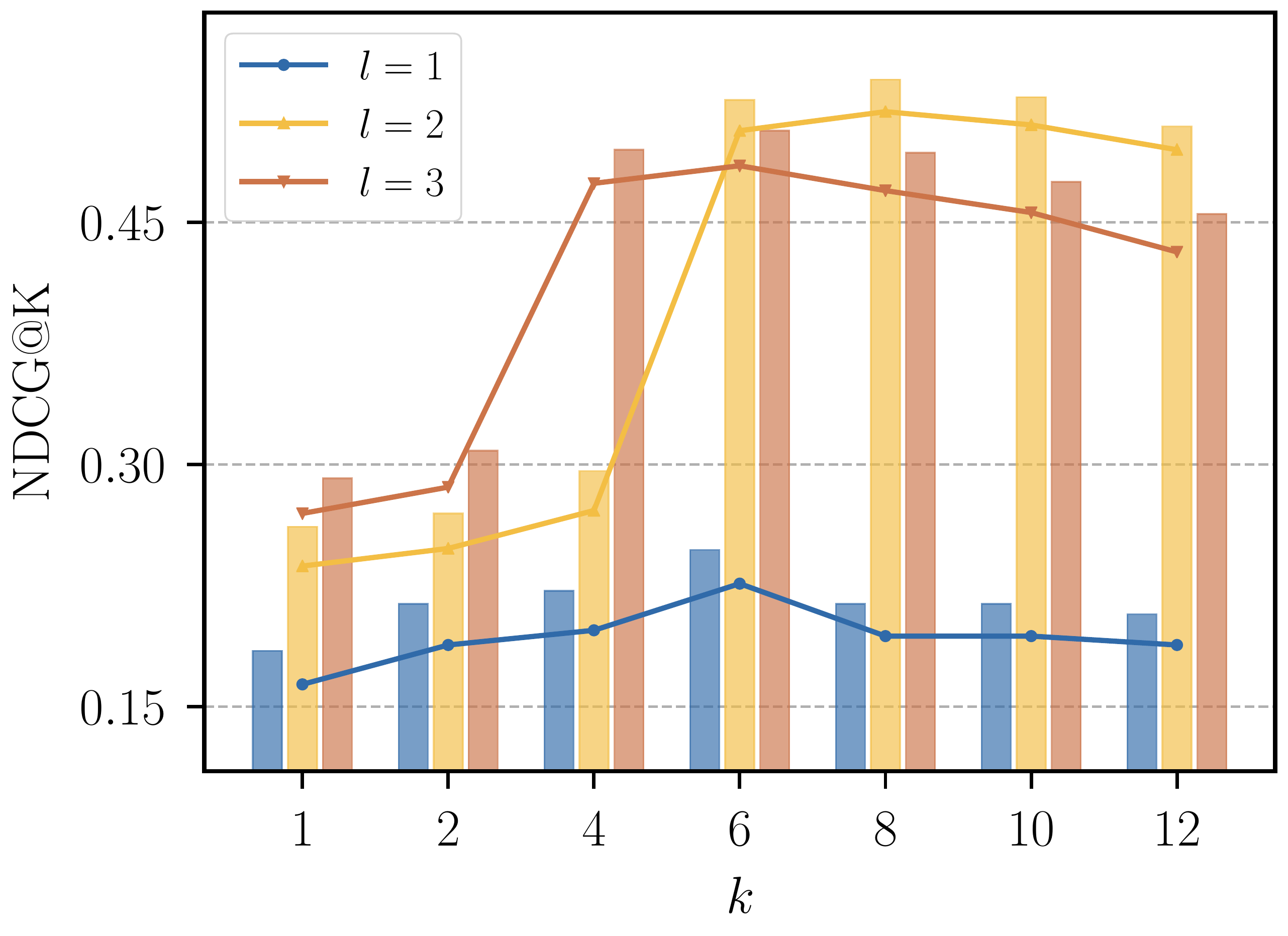}
		\label{fig:param_ndcg_nyc}}
    \end{minipage}
    \begin{minipage}[t]{0.32\linewidth}
        \centering
        \subfigure{
        \includegraphics[width=0.96\linewidth]{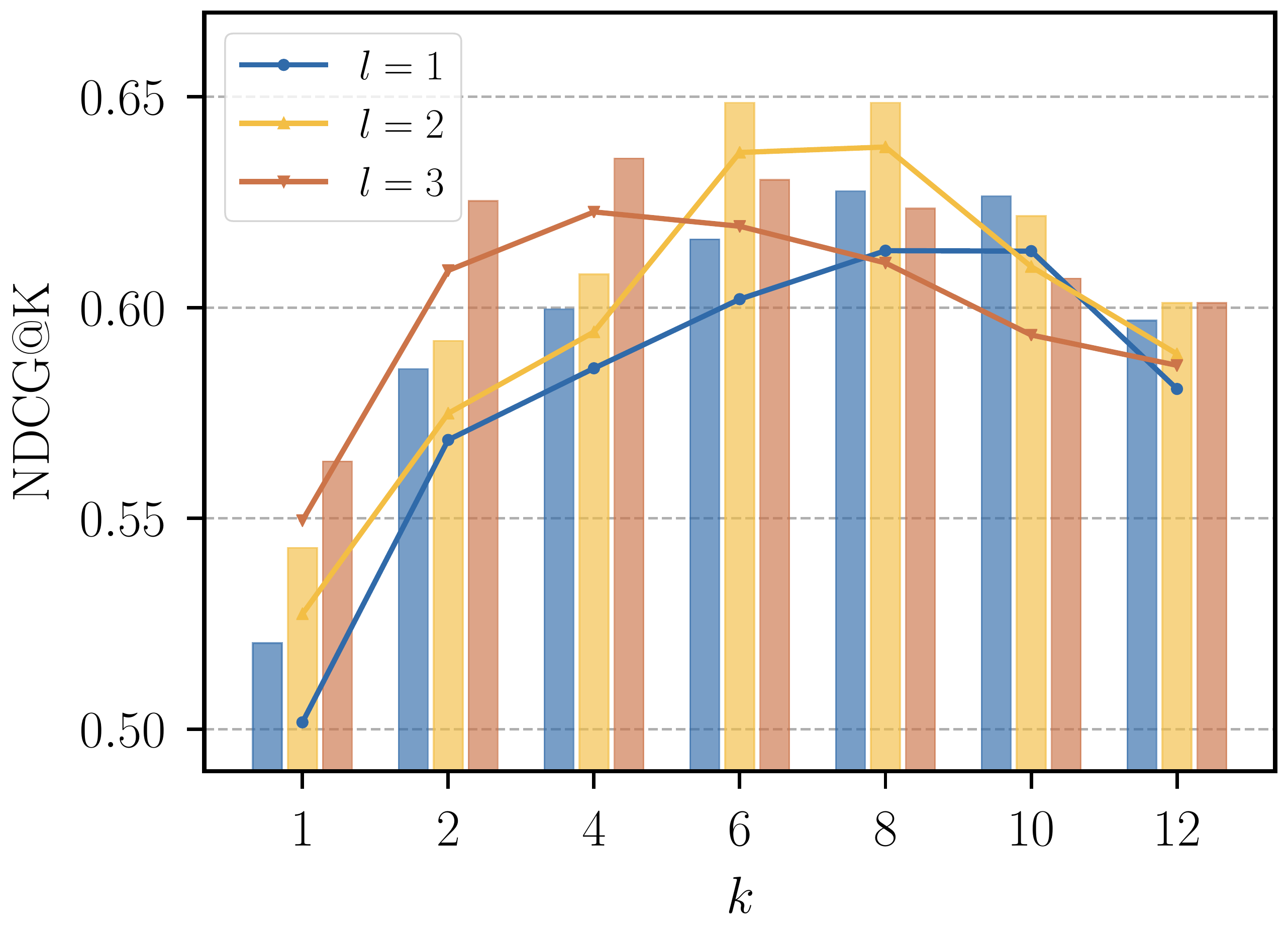}
		\label{fig:param_ndcg_us}}
    \end{minipage}
    \begin{minipage}[t]{0.32\linewidth}
        \centering
        \subfigure{
        \includegraphics[width=0.96\linewidth]{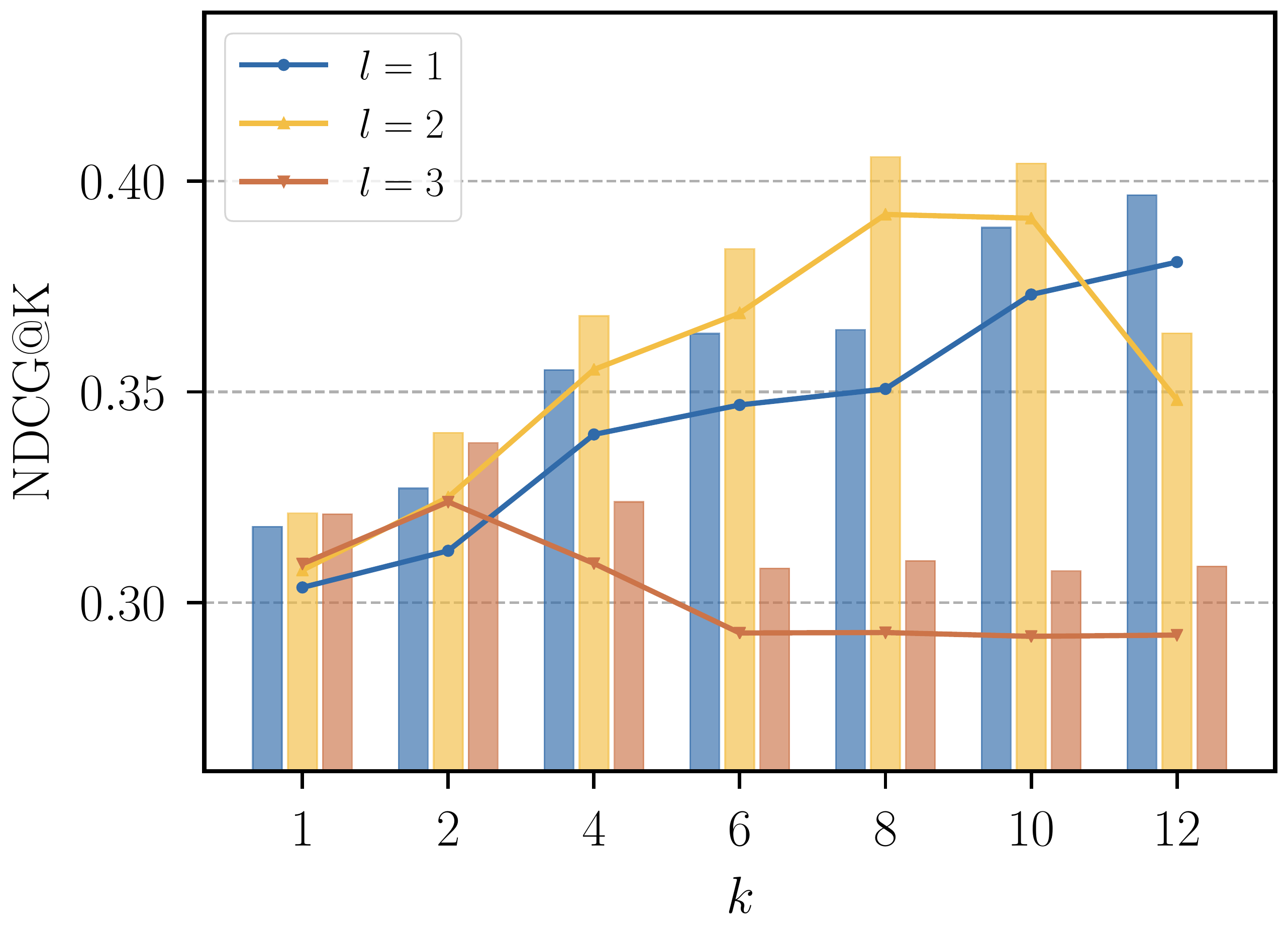}
		\label{fig:param_ndcg_go}}
    \end{minipage}
    \setcounter{subfigure}{0}
    \begin{minipage}[t]{0.32\linewidth}
        \centering
        \subfigure[Foursquare NYC]{
        \includegraphics[width=0.96\linewidth]{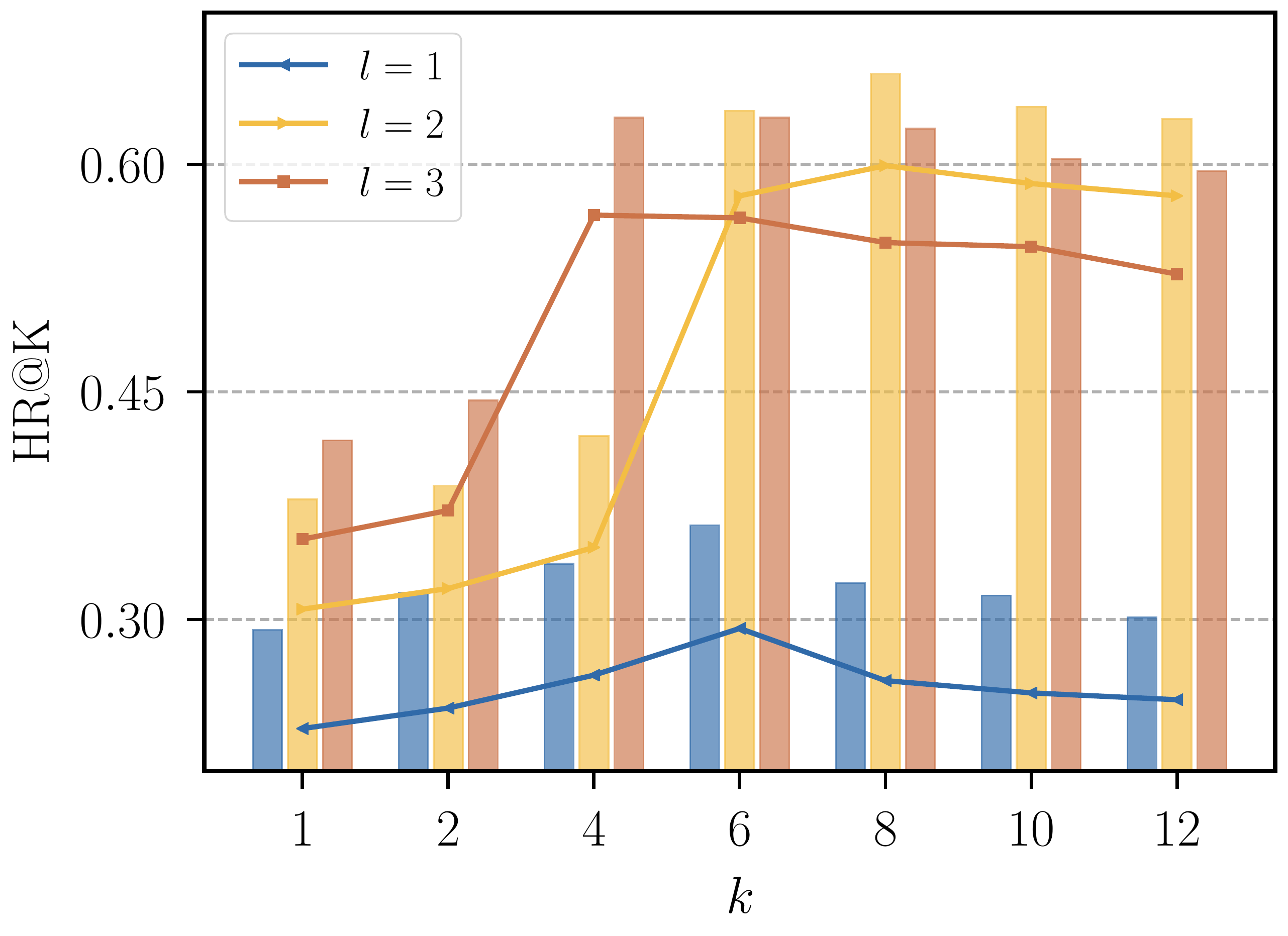}
		\label{fig:param_hr_nyc}}
    \end{minipage}
    \begin{minipage}[t]{0.32\linewidth}
        \centering
        \subfigure[Foursquare US]{
        \includegraphics[width=0.96\linewidth]{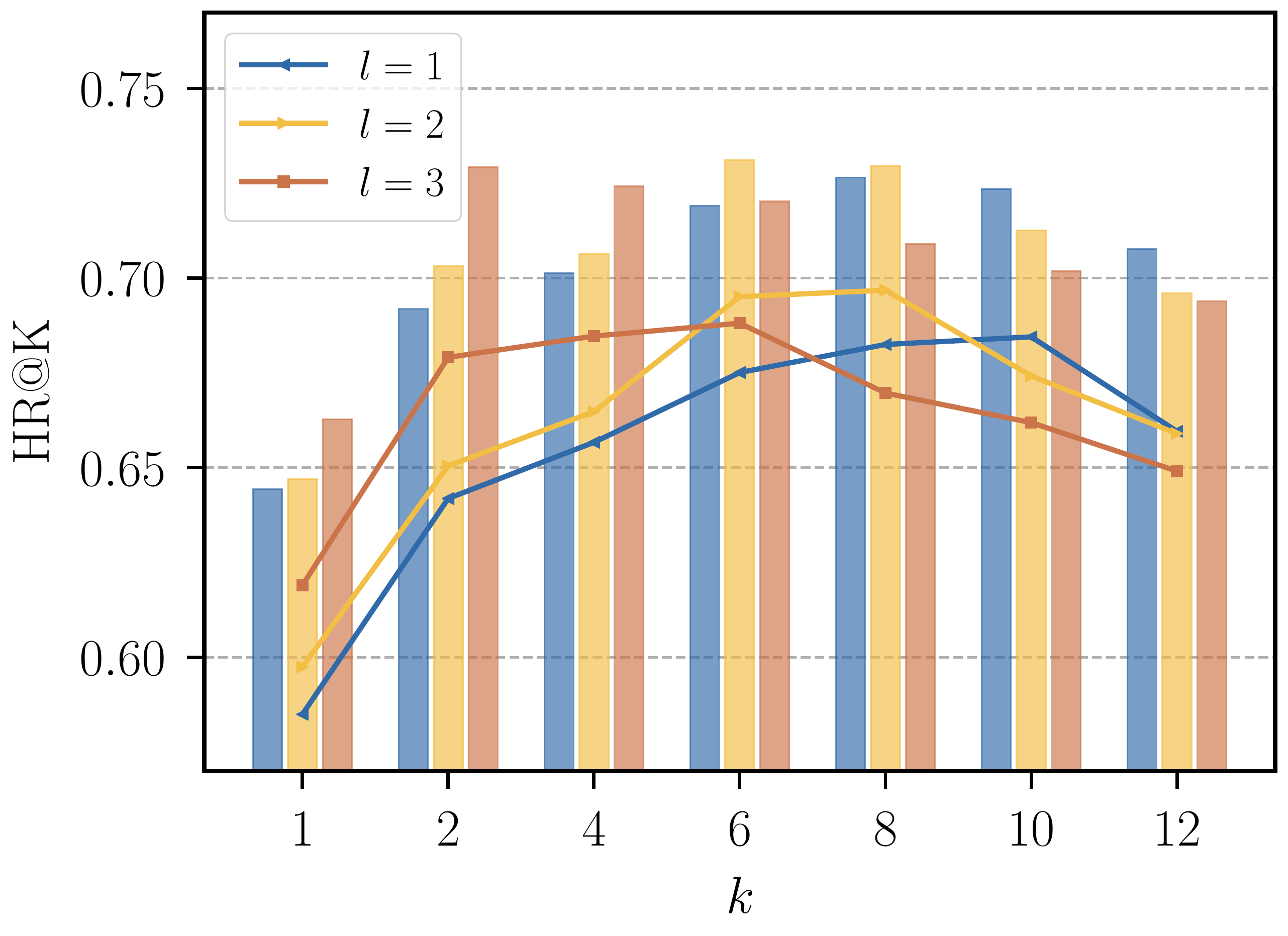}
		\label{fig:param_hr_us}}
    \end{minipage}
    \begin{minipage}[t]{0.32\linewidth}
        \centering
        \subfigure[Gowalla]{
        \includegraphics[width=0.96\linewidth]{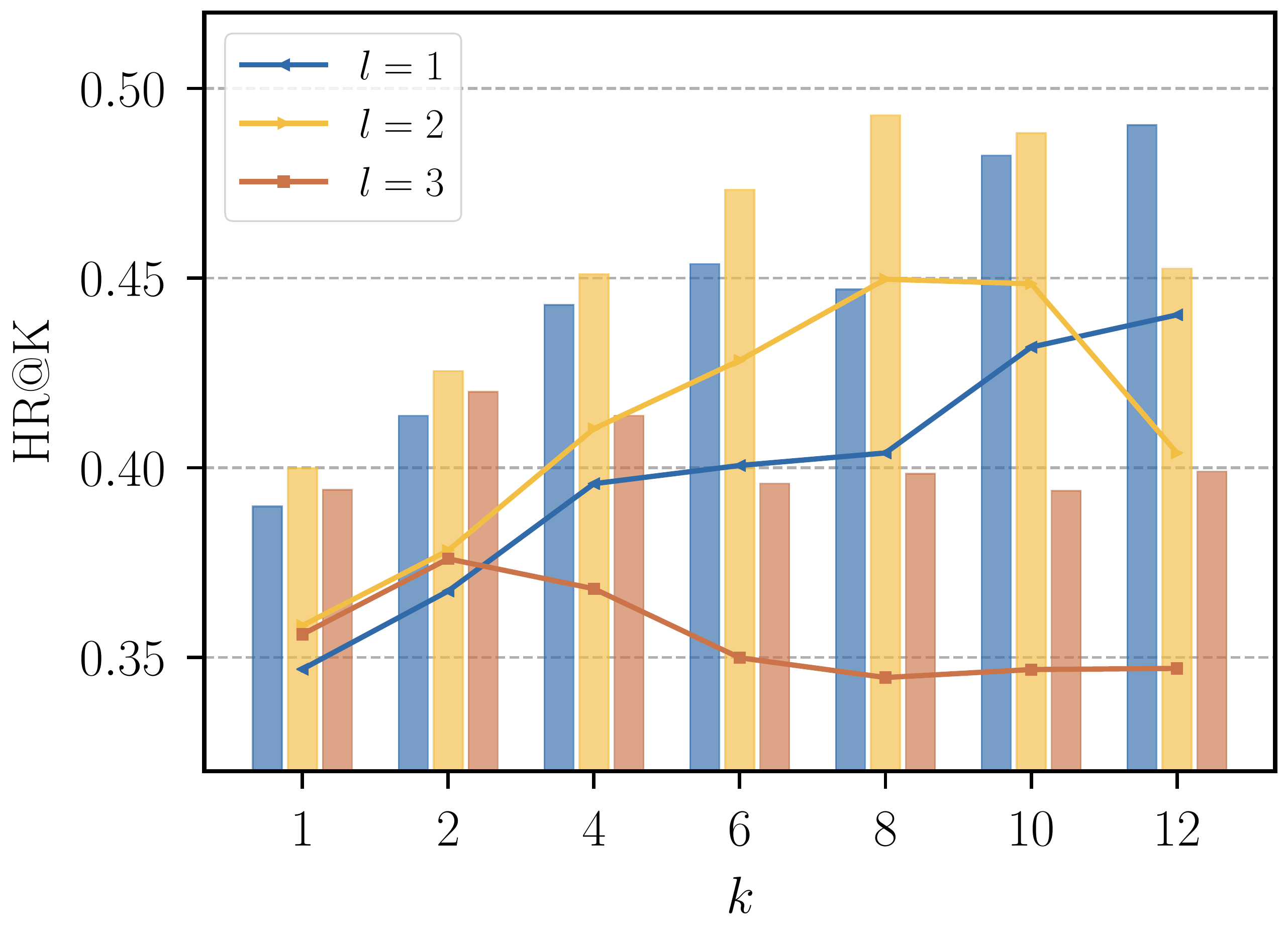}
		\label{fig:param_hr_go}}
    \end{minipage}
\caption{Parameter analysis on three datasets.}
\label{fig:param}
\end{figure*}

The initial length of subsequences and the number of stacked hierarchical encoders play critical roles in STAR-HiT. For example, one-encoder model is unable to capture the hierarchical structure of the check-in sequence, while the semantic subsequences can hardly be discovered if the initial length of subsequences are set too short. On the contrary, too much encoders could lead to an overfitting issue. Besides, if the initial subsequence length is set too long, check-ins with low correlation are aggregated together, which could introduce unwanted noises. Therefore, it is reasonable to assume that there exist optimal settings of these two hyper-parameters for latent hierarchical structure modeling of check-in sequences. Towards this end, we perform experiments with different settings of these two hyper-parameters to investigate their impact on performance. Specifically, we vary the number of hierarchical encoders (\ie, $l$) in $\{1, 2, 3\}$ and the initial subsequence length (\ie, $k$) in $\{1, 2, 4, 6, 8, 10, 12\}$. We use STAR-HiT$_{i:j}$ to denote the STAR-HiT with $i$ hierarchical encoders while the initial subsequence length is set to $j$. The experimental results are summarized in Figure \ref{fig:param}, where lines represent the results \wrtq $K=5$ and bars in same color depict results \wrtq $K=10$ for corresponding settings. From this result, we have the following observations:

\begin{itemize}
    \item STAR-HiT$_{2:8}$ yields the best performance across all the board. Therefore, we set $l=2, k=8$ as default parameters unless otherwise specified. This verifies that STAR-HiT benefits from appropriate settings of $l$ and $k$ to effectively model the latent hierarchical structure of check-in sequences, so as to achieve superior recommendation performance.
    \item With fixed $l$, the performance first increases then drops as $k$ gets larger in most cases. Moreover, with larger $l$, the performance is more likely to peak at smaller $k$. It is also worthwhile to note that on Foursquare NYC and Foursquare US, STAR-HiT$_{2:k}$ performs worse than STAR-HiT$_{3:k}$ when $k<6$, while outperforms STAR-HiT$_{3:k}$ when $k\leq 6$. We attribute these characteristics to the similar role the two parameters play in capturing the multi-granularity semantic subsequences.
    \item STAR-HiT$_{1:k}$ performs worst compared to STAR-HiT$_{2:k}$ or STAR-HiT$_{3:k}$ on Foursquare NYC generally. The reason may lie in the relatively high revisit frequency in Foursquare NYC that exhibits a stronger hierarchical structure of the check-in sequence. 
    \item STAR-HiT$_{3:k}$ falls behind STAR-HiT$_{1:k}$ or STAR-HiT$_{2:k}$ by a large margin in most cases on Gowalla. The possible reasons are two-fold, that is, the low revisit frequency and the high sparsity lead to difficulties in capturing the hierarchical structure of check-in sequences.
\end{itemize}

\begin{figure}[htb]
    \centering
	\begin{minipage}[t]{0.95\linewidth}
		\centering
		\subfigure[$l=1$]{
		\includegraphics[width=0.99\linewidth]{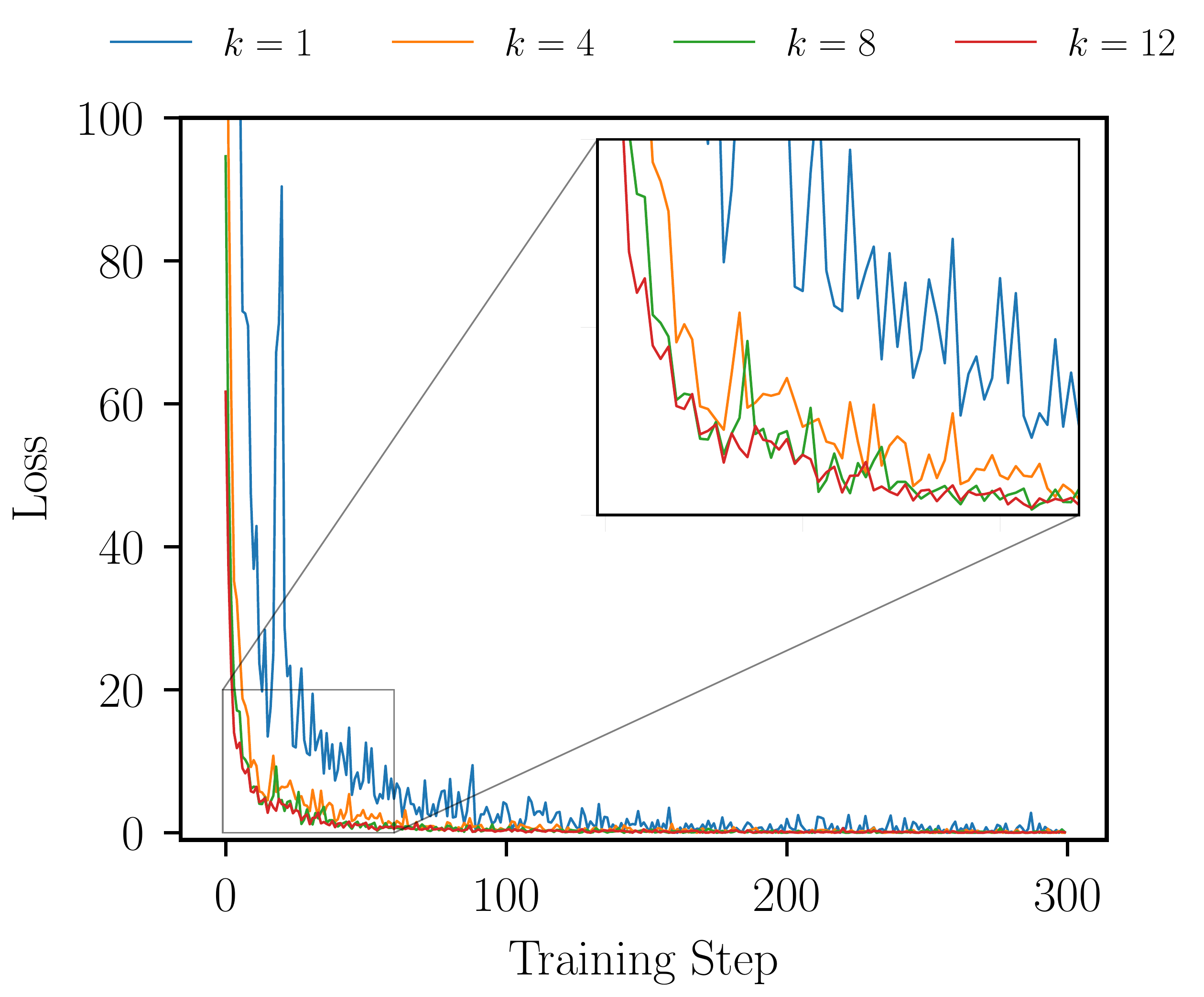}
		\label{fig:nyc_train_l1}}
	\end{minipage}
	\begin{minipage}[t]{0.95\linewidth}
		\centering
		\subfigure[$l=2$]{
		\includegraphics[width=0.99\linewidth]{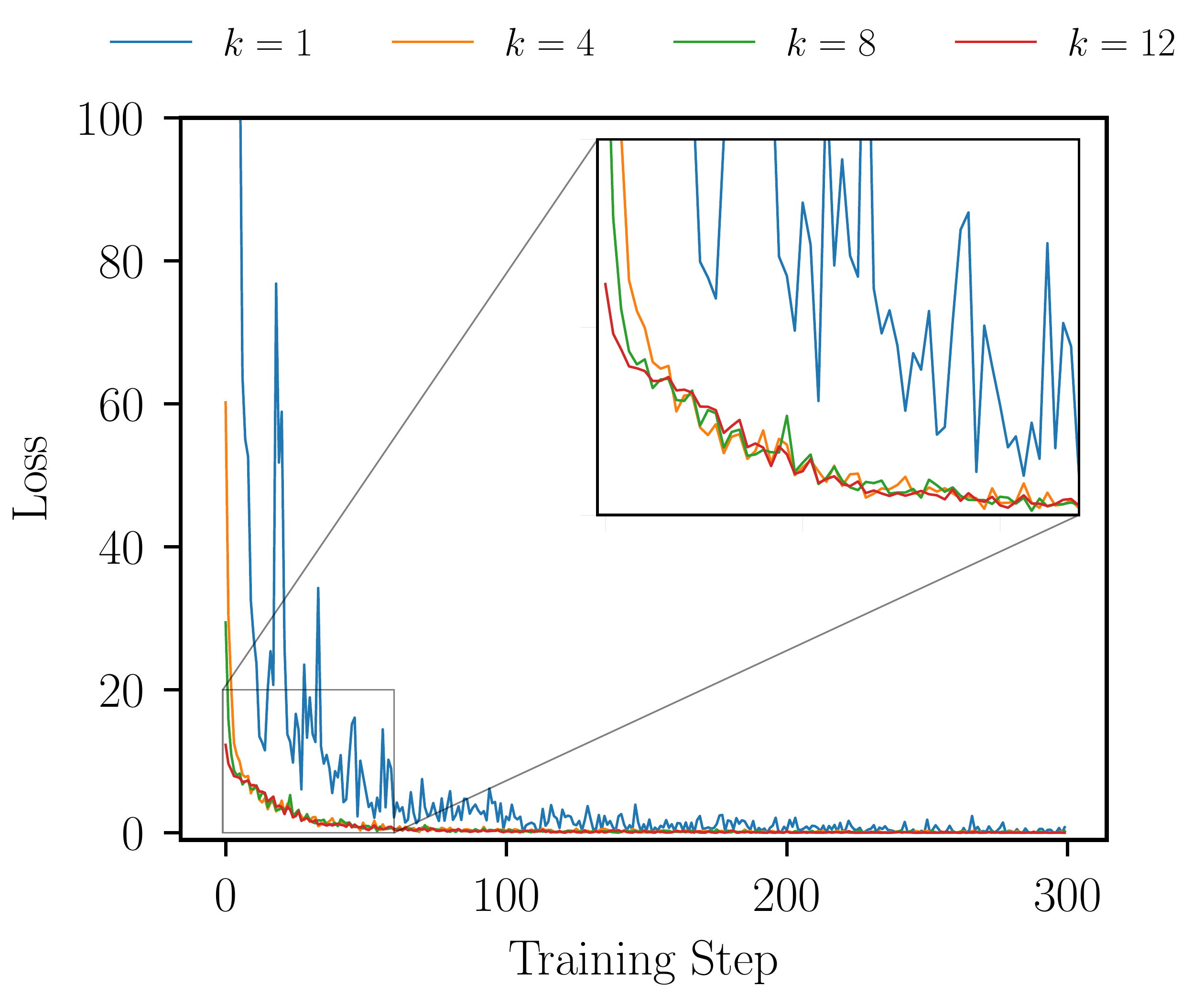}
		\label{fig:nyc_train_l2}}
	\end{minipage}
\caption{Training efficiency on Foursquare NYC with different initial subsequence lengths $k$.}
\label{fig:training}
\end{figure}

Taking Foursquare NYC as an example, we analyze the training efficiency of STAR-HiT with different settings of $l$ and $k$. The training loss curves are illustrated in Figure \ref{fig:training}. As we can see, STAR-HiT$_{1:1}$ and STAR-HiT$_{2:1}$ exhibit large fluctuations compared to models with higher $k$. When $l=1$, the training loss curve of STAR-HiT$_{1:4}$ is generally lower and more stable than STAR-HiT$_{1:8}$ and STAR-HiT$_{1:12}$, while such differences are unnoticeable when $l=2$. In addition, when $k>1$, the loss of STAR-HiT$_{2:k}$ drops faster and more steadily to nearly zero compared to STAR-HiT$_{1:k}$. Jointly analyzing Figure \ref{fig:param} and Figure \ref{fig:training}, we again verify that with the suitable settings of $l$ and $k$, STAR-HiT is capable of discovering the semantic subsequences in check-in sequences, thereby better uncovering the hierarchical structure present in user sequential behavior patterns.

\subsubsection{Effect of the embedding dimension}

\begin{figure*}[htb]
    \centering
    \begin{minipage}[t]{0.3\linewidth}
        \centering
        \subfigure{
        \includegraphics[width=0.96\linewidth]{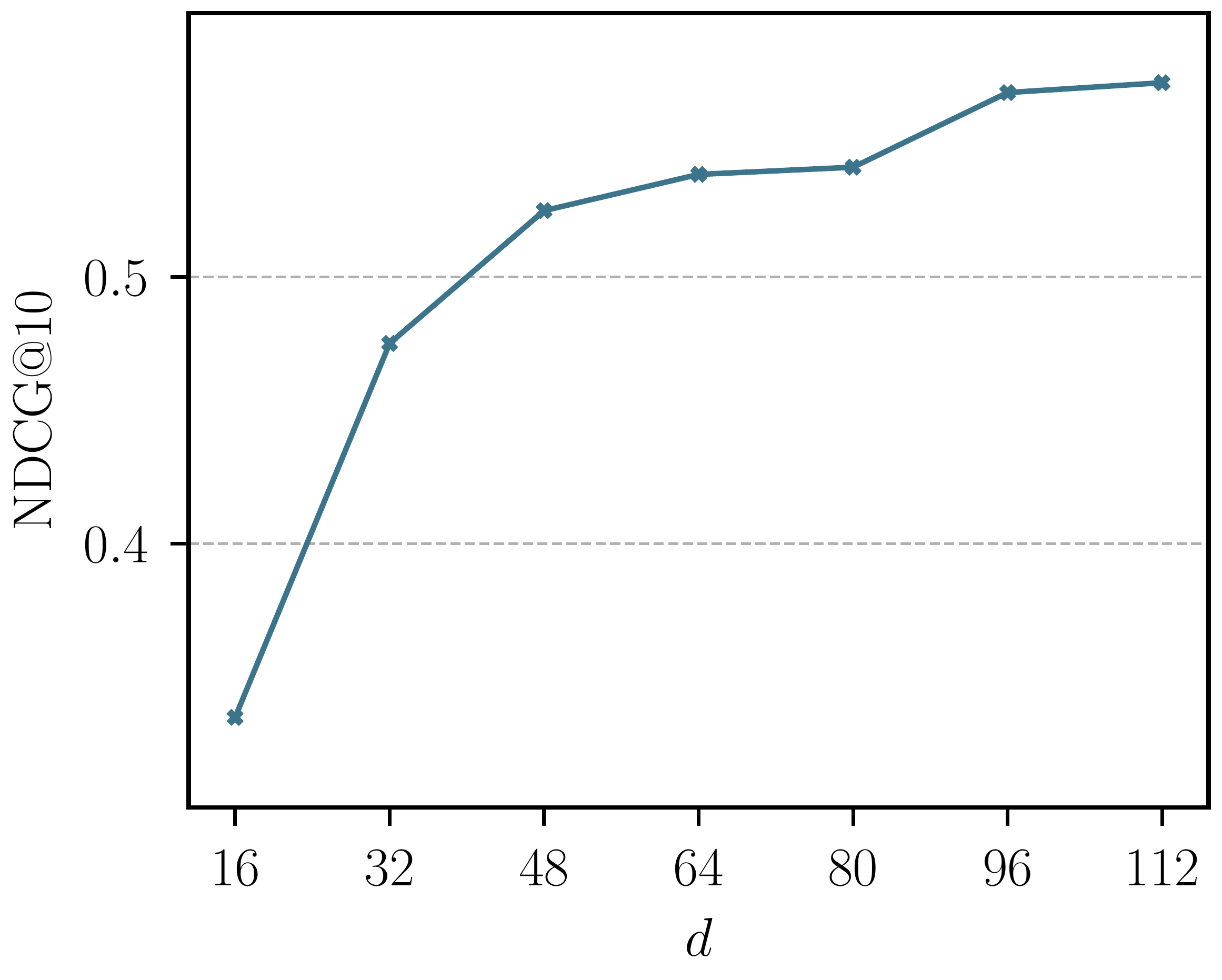}}
    \end{minipage}
    \begin{minipage}[t]{0.3\linewidth}
        \centering
        \subfigure{
        \includegraphics[width=0.96\linewidth]{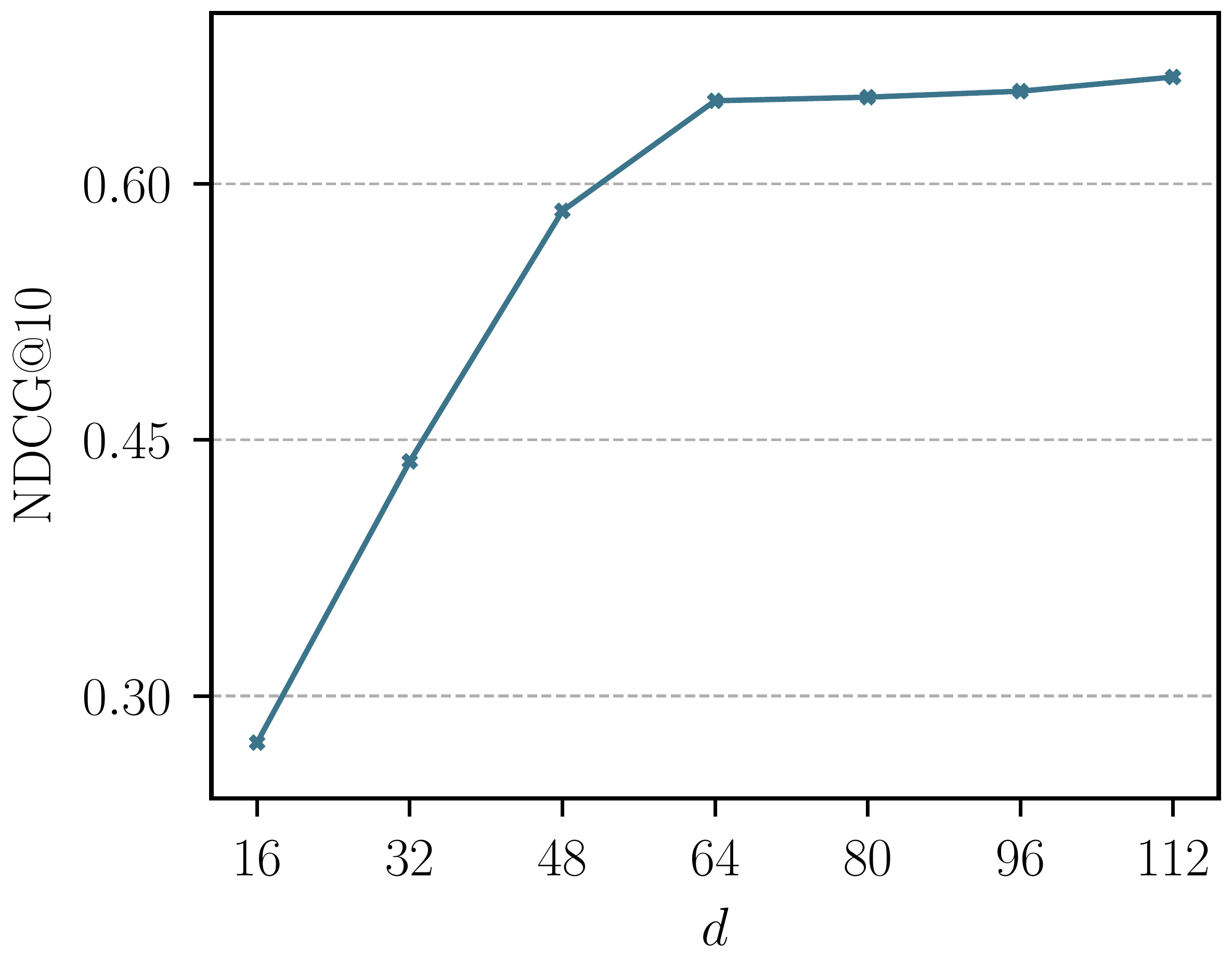}}
    \end{minipage}
    \begin{minipage}[t]{0.3\linewidth}
        \centering
        \subfigure{
        \includegraphics[width=0.96\linewidth]{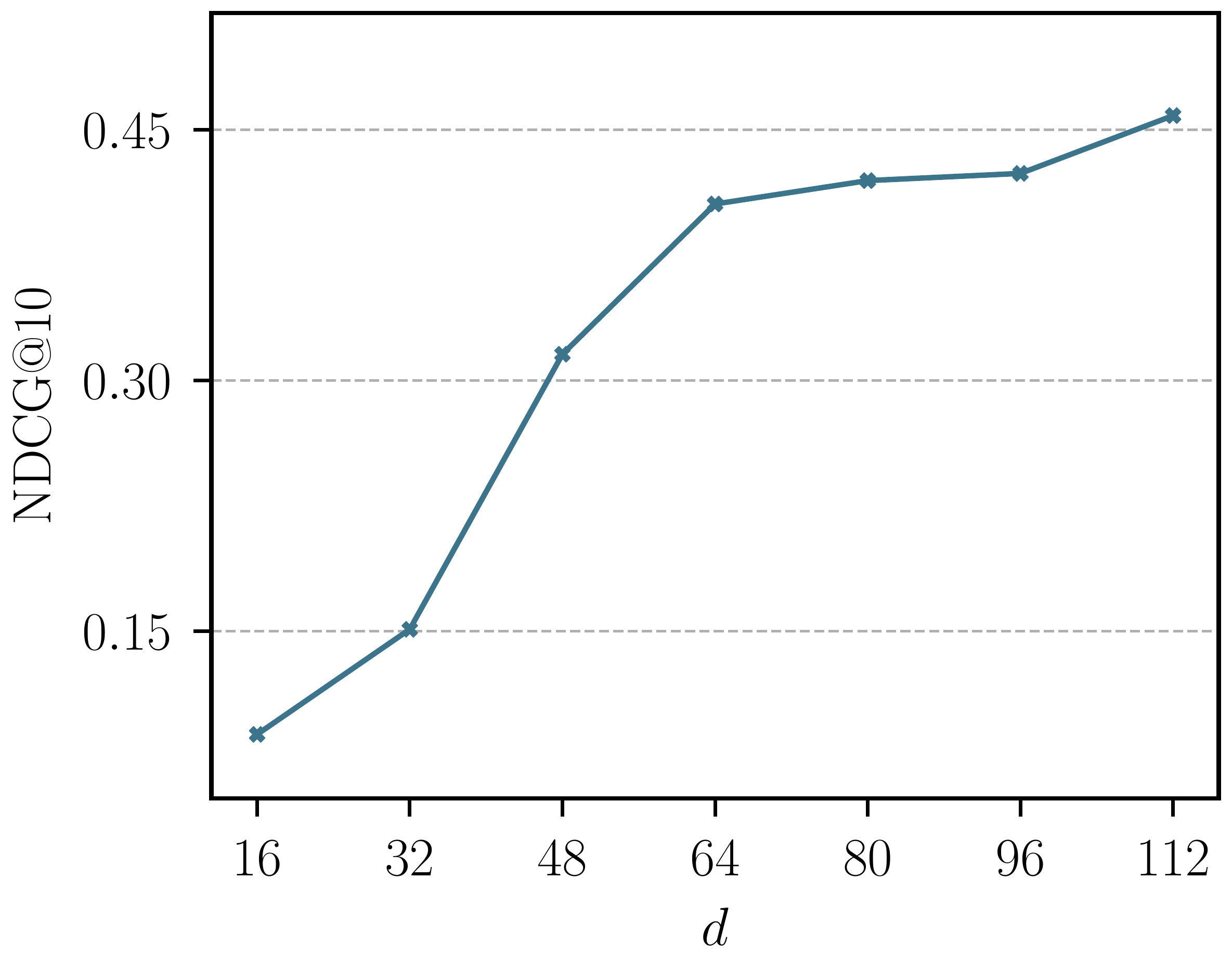}}
    \end{minipage}
    \setcounter{subfigure}{0}
    \begin{minipage}[t]{0.3\linewidth}
        \centering
        \subfigure[Foursquare NYC]{
        \includegraphics[width=0.96\linewidth]{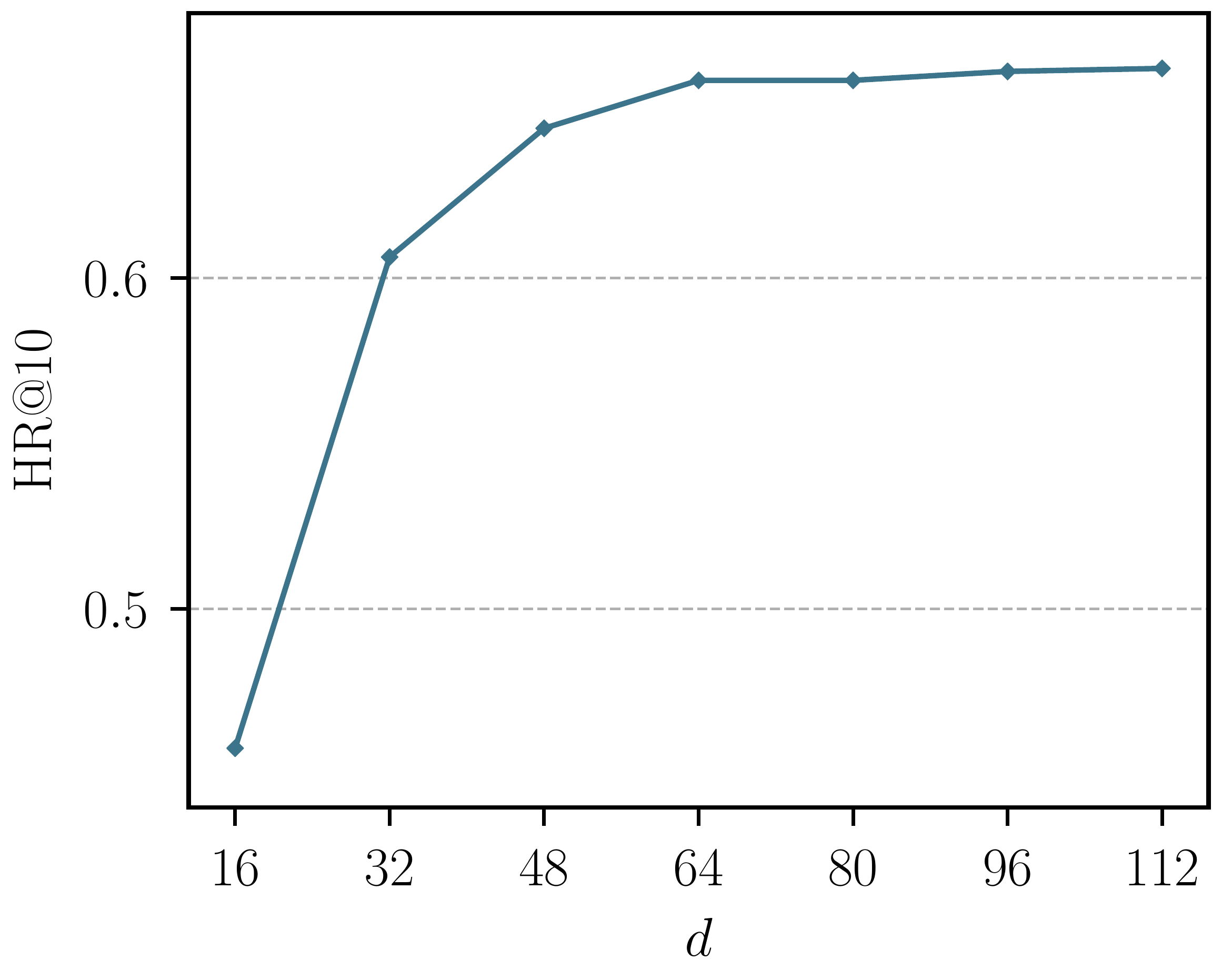}}
    \end{minipage}
    \begin{minipage}[t]{0.3\linewidth}
        \centering
        \subfigure[Foursquare US]{
        \includegraphics[width=0.96\linewidth]{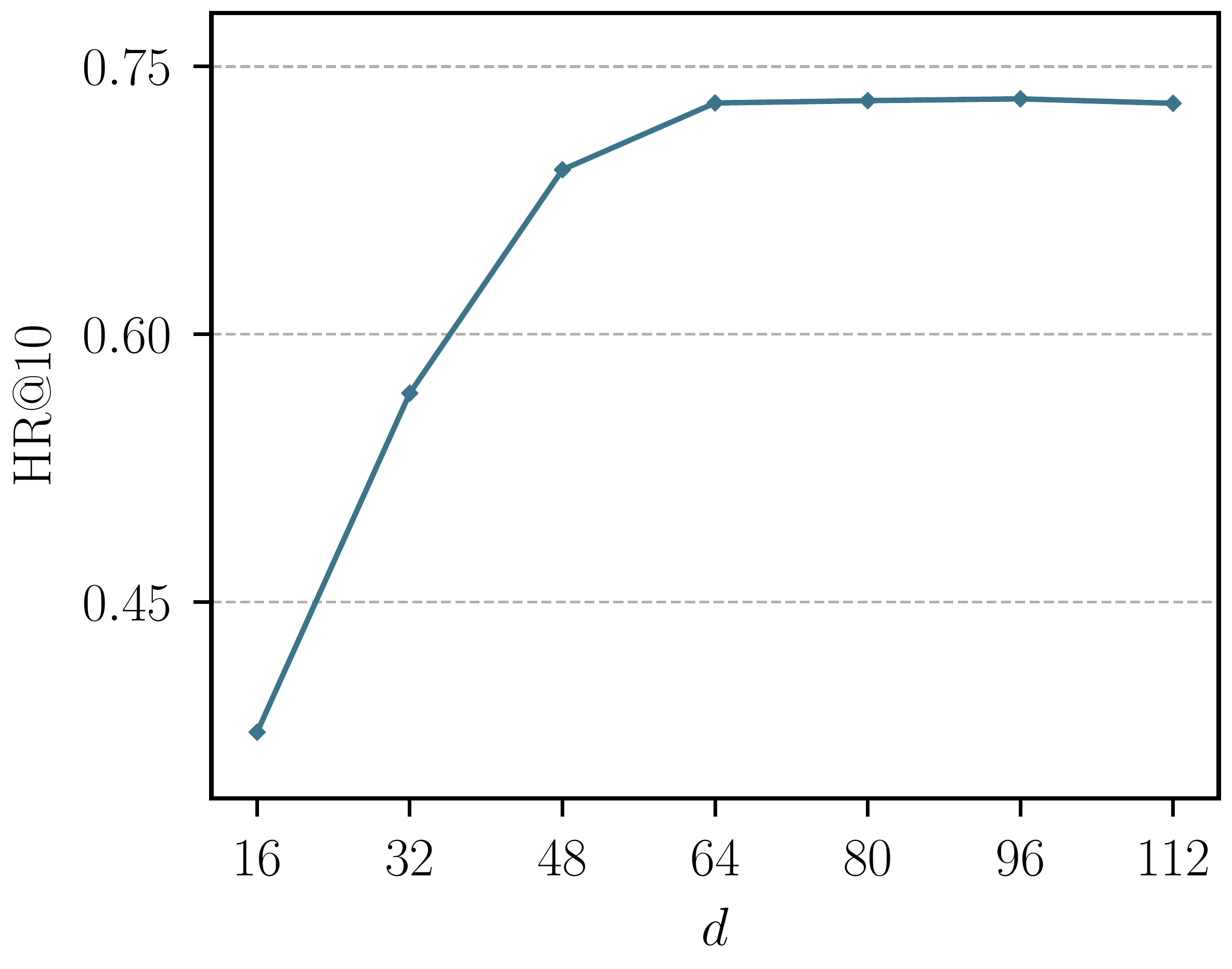}}
    \end{minipage}
    \begin{minipage}[t]{0.3\linewidth}
        \centering
        \subfigure[Gowalla]{
        \includegraphics[width=0.96\linewidth]{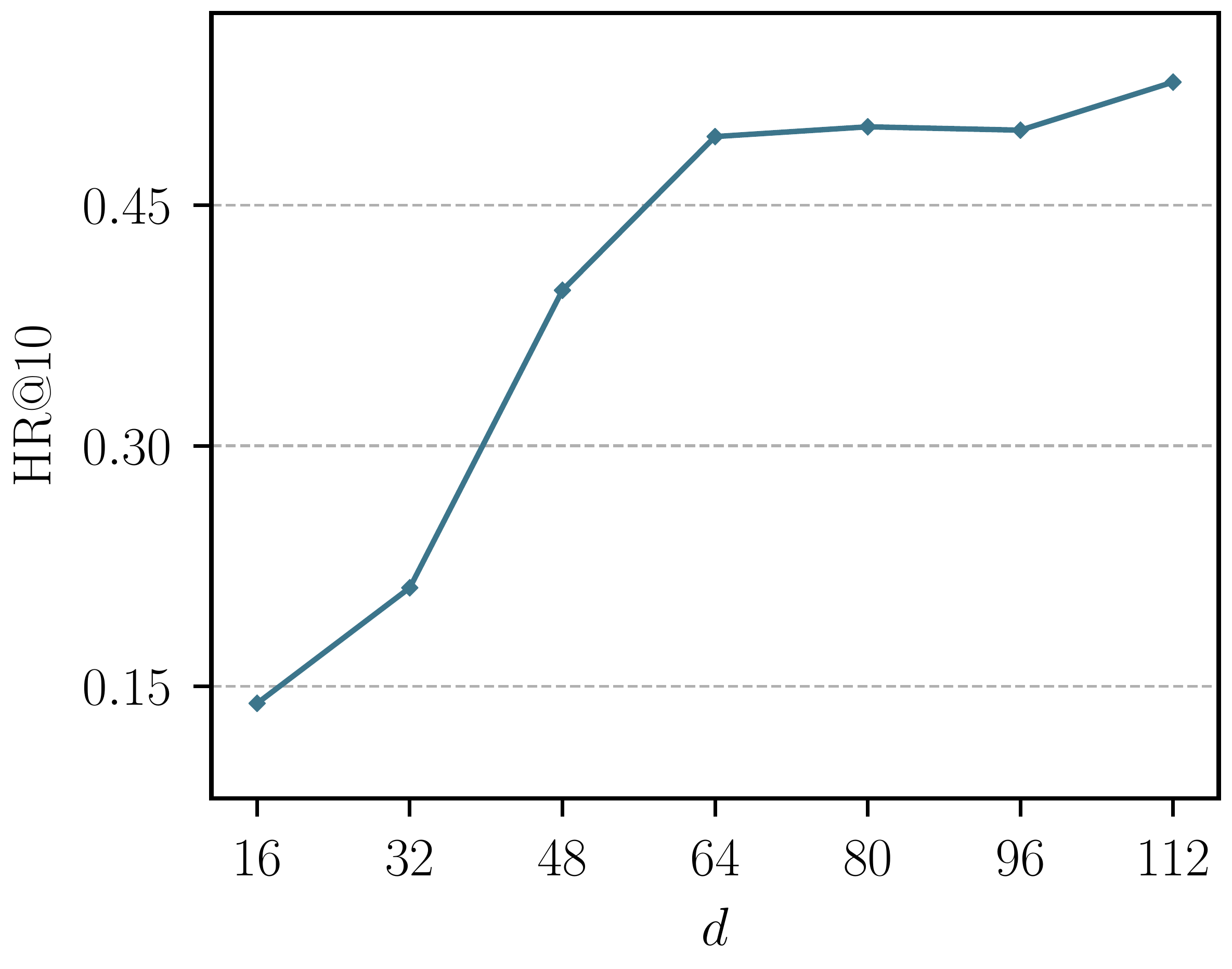}}
    \end{minipage}
\caption{Effect of the embedding dimension $d$.}
\label{fig:dim}
\end{figure*}

We also conduct experiments to analyze the effect of the embedding dimension (\ie, $d$) used in STAR-HiT. In particular, we set the embedding dimension from 16 to 112, with a step of 16. Figure \ref{fig:dim} illustrates the experimental results in terms of NDCG@10 and HR@10 on three datasets. From Figure \ref{fig:dim}, we can see that the performance gets much worse when using a small embedding dimension, as it is insufficient to encode the contextual information. As the embedding dimension grows, the performance first increases dramatically and then gradually becomes stable. Considering the trade-off between cost and performance, we set the default embedding dimension $d$ to 64.

\subsubsection{Ablation Study}

As stated, there are two essential components of the proposed hierarchical encoder, including: 1) \textbf{attention module}, including the global attention layer and local attention layer, which captures global context and local context to enhance representations; 2) \textbf{subsequence aggregation module}, including the sequence partition layer and subsequence aggregation layer, which utilizes contextual information to reason and locate subsequences, thereby obtaining semantic subsequences. To analyze the effectiveness of the various components, we conduct an ablation study by considering the following variants:

\begin{itemize}
    \item STAR-HiT$_{\operatorname{GA}}$: The global attention layer in each hierarchical encoder is removed, which captures the global spatio-temporal context to enhance representations.
    \item STAR-HiT$_{\operatorname{LA}}$: The local attention layer in each hierarchical encoder is removed, which injects the local context of the extracted subsequence into each representations within.
    \item STAR-HiT$_{\operatorname{LGA}}$: Both the global attention layer and local attention layer are removed, such that the model solely hierarchically partitions the sequence and aggregates the subsequences without any contextual enhancement.
    \item STAR-HiT$_{\operatorname{AGG2}}$: The hierarchical structure modeling through the sequence partition layer and subsequence aggregation layer is disabled. The local attention layer is removed as well, since it is calculated within the identified subsequence. Overall, the proposed hierarchical encoder degrades to the vanilla Transformer encoder that solely learns from POI-to-POI interactions.
    \item STAR-HiT$_{\operatorname{AGG4}}$: Since there are two attention layers in a hierarchical encoder, we also double the number of encoders (\ie, $l=4$) for STAR-HiT$_{\operatorname{AGG2}}$, so as to involve the same number of attention layers as the original STAR-HiT.
\end{itemize}

\begin{table*}[htb]
\centering
\caption{Performance Comparison (NDCG@K) with STAR-HiT Variants}
\begin{tabular}{lcccccc}
\toprule
& \multicolumn{2}{c}{Foursquare NYC} & \multicolumn{2}{c}{Foursquare US} & \multicolumn{2}{c}{Gowalla} \\ \cmidrule(lr){2-3} \cmidrule(lr){4-5} \cmidrule(lr){6-7}
& NDCG@5 & NDCG@10 & NDCG@5 & NDCG@10 & NDCG@5 & NDCG@10 \\ \midrule
STAR-HiT$_{\operatorname{GA}}$ & 0.3430 & 0.3602 & 0.5612 & 0.5756 & 0.2988 & 0.3162 \\ 
STAR-HiT$_{\operatorname{LA}}$ & 0.4715 & 0.4914 & 0.6213 & 0.6341 & 0.3569 & 0.3728 \\ 
STAR-HiT$_{\operatorname{GLA}}$ & 0.3059 & 0.3158 & 0.5359 & 0.5505 & 0.2825 & 0.2970 \\
STAR-HiT$_{\operatorname{AGG2}}$ & 0.1919 & 0.2154 & 0.5457 & 0.5608 & 0.3459 & 0.3621 \\
STAR-HiT$_{\operatorname{AGG4}}$ & 0.2136 & 0.2326 & 0.3540 & 0.3729 & 0.2556 & 0.2651 \\ \midrule
STAR-HiT & 0.5186 & 0.5385 & 0.6381 & 0.6486 & 0.3921 & 0.4057 \\ \bottomrule
\end{tabular}
\label{tab:perf_variants}
\end{table*}

Table \ref{tab:perf_variants} compares the performance of STAR-HiT and its variants in terms of NDCG@5 and NDCG@10. From Table \ref{tab:perf_variants}, we have three key observations:

\begin{itemize}
    \item \textbf{Attention module: }Removing the global attention layer leads to significant performance drops in terms of NDCG@5 by 12.05\%-33.86\% and NDCG@10 by 11.26\%-33.11\% on three datasets. Besides, removing the local attention layer leads to relatively slight performance drops with regard to NDCG@5 by 2.63\%-9.08\% and NDCG@10 by 2.23\%-8.74\%. Since the local attention is calculated within each extracted subsequence, it may be affected by the performance of the subsequence partition layer. This may be the reason why the local attention layer provides less performance improvement compared to the global attention layer. Overall, jointly removing the attention module makes the model unable to make full use of the context, resulting in even worse performance, that is, the performance drops in terms of NDCG@5 by 16.01\%-41.01\% and NDCG@10 by 15.11\%-41.34\%.
    \item \textbf{Subsequence aggregation module: }Disabling the multi-granularity subsequence aggregation, STAR-HiT$_{\operatorname{AGG2}}$ performs 11.78\%-63.00\%, 10.75\%-60.00\% worse than STAR-HiT in terms of NDCG@5, NDCG@10, respectively. STAR-HiT$_{\operatorname{AGG4}}$ yields worse performance as well. Surprisingly, STAR-HiT$_{\operatorname{AGG4}}$ outperforms STAR-HiT$_{\operatorname{AGG2}}$ on Foursquare NYC, while performs notably worse on others. The results indicate that STAR-HiT$_{\operatorname{AGG4}}$ may suffer from overfitting issue due to the sparsity of data. The performance comparison demonstrates the significance of involving not only POI-to-POI interactions but also subsequence-level context and interactions for recommendations, which is achieved by STAR-HiT.
    \item As expected, STAR-HiT outperforms all the variants by a large margin. Moreover, jointly analyzing Table \ref{tab:perf_baselines} and Table \ref{tab:perf_variants}, we can see that the variants of STAR-HiT still achieve competitive performance compared to baseline methods. The results emphasize the significance of joint context modeling through attention mechanisms and hierarchical structure modeling through multi-granularity semantic subsequence discovering.
\end{itemize}

\begin{table*}[htb]
\centering
\caption{Performance Comparison (NDCG@K) with STAR-HiT Variants of Fixed-Length Subsequences}
\begin{tabular}{lcccccc}
\toprule
& \multicolumn{2}{c}{Foursquare NYC} & \multicolumn{2}{c}{Foursquare US} & \multicolumn{2}{c}{Gowalla} \\ \cmidrule(lr){2-3} \cmidrule(lr){4-5} \cmidrule(lr){6-7}
& NDCG@5 & NDCG@10 & NDCG@5 & NDCG@10 & NDCG@5 & NDCG@10 \\ \midrule
STAR-HiT$_{\operatorname{GA}}$ & 0.3430 & 0.3602 & 0.5612 & 0.5756 & 0.2988 & 0.3162 \\ 
STAR-HiT$_{\operatorname{GA-F}}$ & 0.1511 & 0.1828 & 0.5296 & 0.5487 & 0.2719 & 0.2860 \\ \midrule
STAR-HiT$_{\operatorname{LA}}$ & 0.4715 & 0.4914 & 0.6213 & 0.6341 & 0.3569 & 0.3728 \\ 
STAR-HiT$_{\operatorname{LA-F}}$ & 0.2860 & 0.3050 & 0.5439 & 0.5635 & 0.3405 & 0.3556 \\ \midrule
STAR-HiT$_{\operatorname{GLA}}$ & 0.3059 & 0.3158 & 0.5359 & 0.5505 & 0.2825 & 0.2970 \\
STAR-HiT$_{\operatorname{GLA-F}}$ & 0.1430 & 0.1665 & 0.5085 & 0.5262 & 0.2612 & 0.2758 \\ \midrule
STAR-HiT & 0.5186 & 0.5385 & 0.6381 & 0.6486 & 0.3921 & 0.4057 \\
STAR-HiT$_{\operatorname{F}}$ & 0.3014 & 0.3240 & 0.5686 & 0.5874 & 0.3247 & 0.3411 \\ \bottomrule
\end{tabular}
\label{tab:perf_variants_2}
\end{table*}

The sequence partition layer in the hierarchical encoder adaptively partitions the input sequence into multiple semantic subsequences, wherein the offset (\ie, ${dx}_i$) and length (\ie, $k_i$) for each subsequence are learned. Offsets are predicted to shift the subsequences towards check-ins with strong correlation, whereas lengths are utilized to better maintain local semantics. They are both supposed to facilitate the performance of STAR-HiT. 

Accordingly, we conduct experiments to explore how the learnable subsequence locations affect the performance. In particular, we implement the variants of STAR-HiT, STAR-HiT$_{\operatorname{GA}}$, STAR-HiT$_{\operatorname{LA}}$, STAR-HiT$_{\operatorname{GLA}}$ by disabling the learnable offsets and lengths, that is, subsequences are fixed to be extracted by uniformly partitioning the input sequence. We denote these variants as STAR-HiT$_{\operatorname{F}}$, STAR-HiT$_{\operatorname{GA-F}}$, STAR-HiT$_{\operatorname{LA-F}}$, and STAR-HiT$_{\operatorname{GLA-F}}$, respectively. The experimental results in terms of NDCG@5 and NDCG@10 are shown in Table \ref{tab:perf_variants_2}. In general, models with learnable locations of subsequences improve over models with fixed-length subsequence extraction by 64.87\%-126.95\%, 5.40\%-14.24\%, 4.82\%-9.91\% in terms of NDCG@5, and 89.67\%-97.03\%, 4.63\%-12.52\%, 4.82\%-10.57\% in terms of NDCG@10 on three datasets. It is worth noting that learnable locations of subsequence boost the performance substantially in Foursquare NYC. This may be due to the stronger hierarchical structure present in check-in sequences in Foursquare NYC, which is consistent with the findings in Section \ref{sec:params}. Partitioning the sequence into subsequences in a fixed way could potentially destroy the semantic information, eventually resulting in the failure to capture the hierarchical structure in check-in sequences. By adaptively partition the sequence into subsequences with different positions and lengths learned from contextual information, STAR-HiT is able to model the hierarchical structure of the check-in sequence by well preserving the semantics in subsequences.

\subsection{Case Study (RQ3)}

Jointly modeling the spatio-temporal context and multi-granularity semantic subsequences, STAR-HiT makes full use of the personalized sequential behavior pattern to predict the next visiting POI. In order to better understand the capability of STAR-HiT to model the latent hierarchical structure, we randomly select a check-in sequence sample (\textit{uid}: 172) on the test set from Foursquare NYC to conduct a case study.

\begin{figure}[htb]
    \centering
    \subfigure[E-1, H-1]{
        \includegraphics[height=0.21\linewidth]{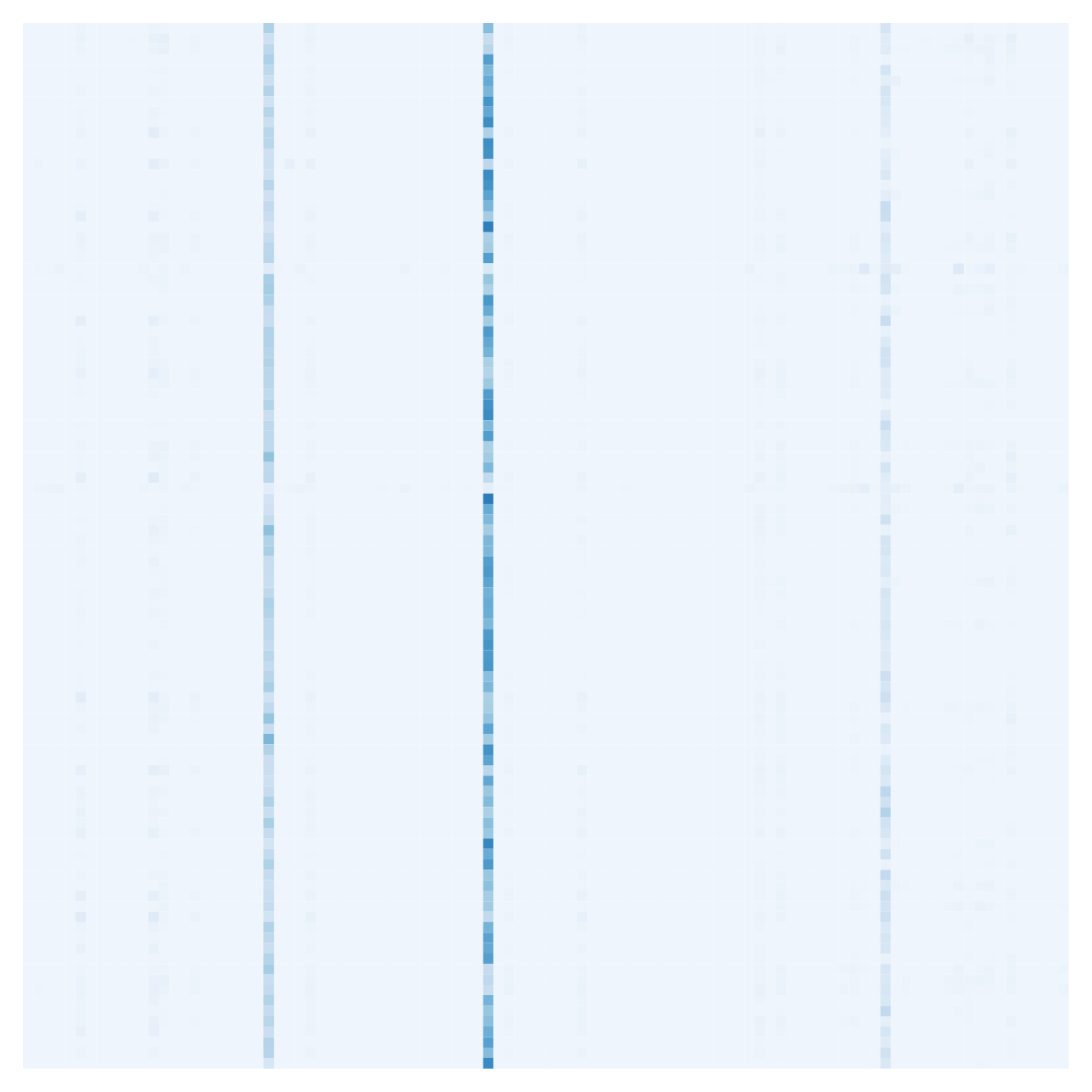}}
    \subfigure[E-1, H-2]{
        \includegraphics[height=0.21\linewidth]{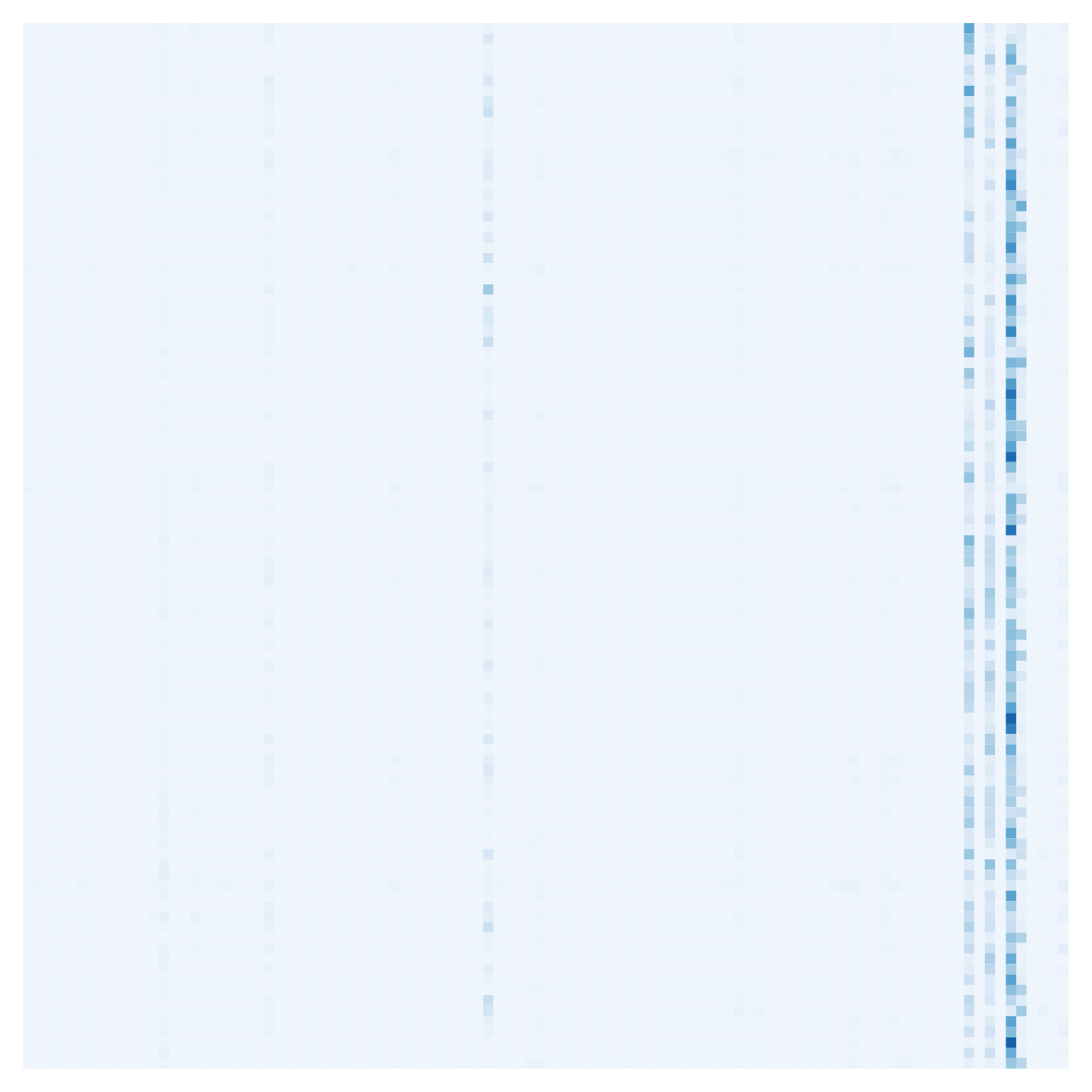}}
    \subfigure[E-1, H-3]{
        \includegraphics[height=0.21\linewidth]{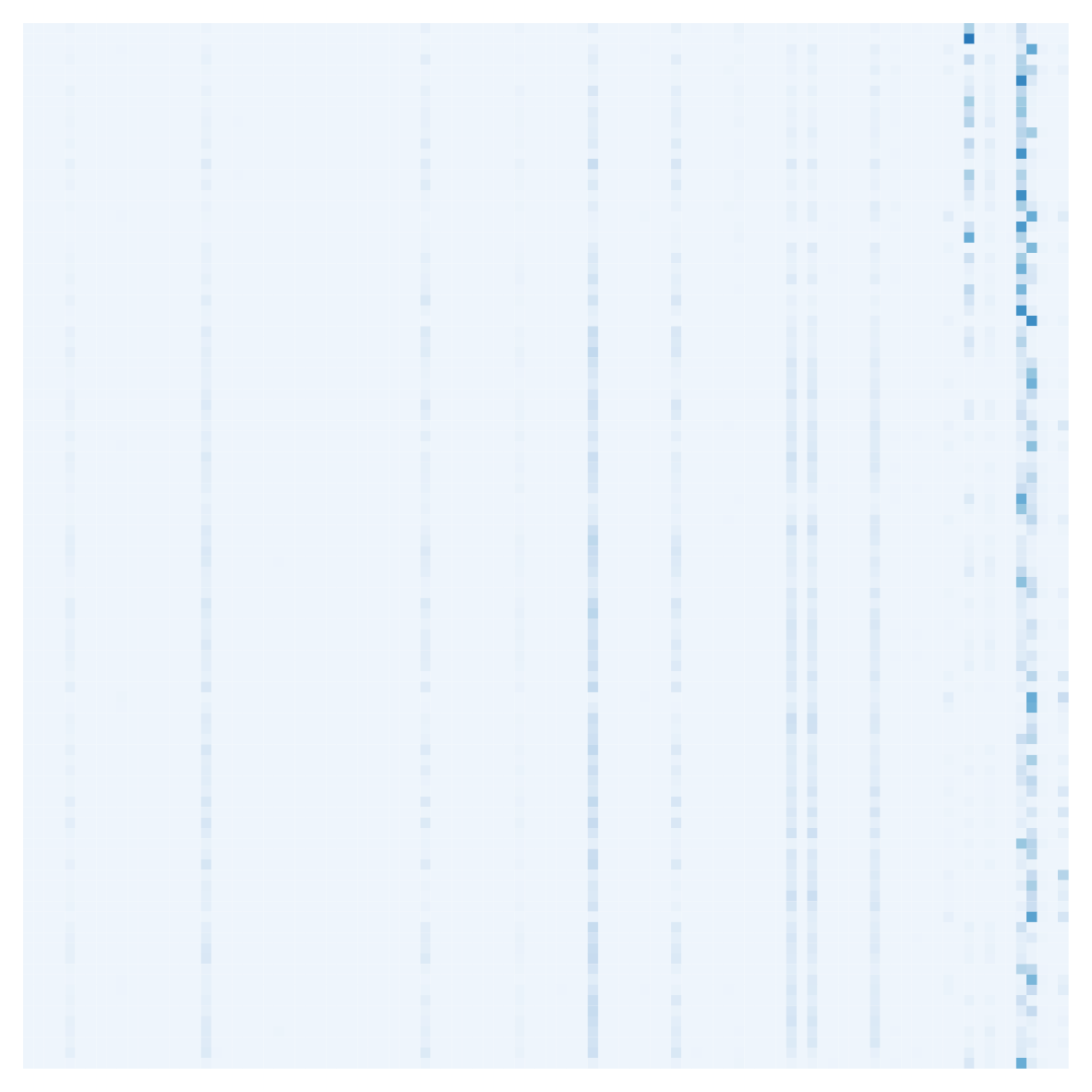}}
    \subfigure[E-1, H-4]{
        \includegraphics[height=0.21\linewidth]{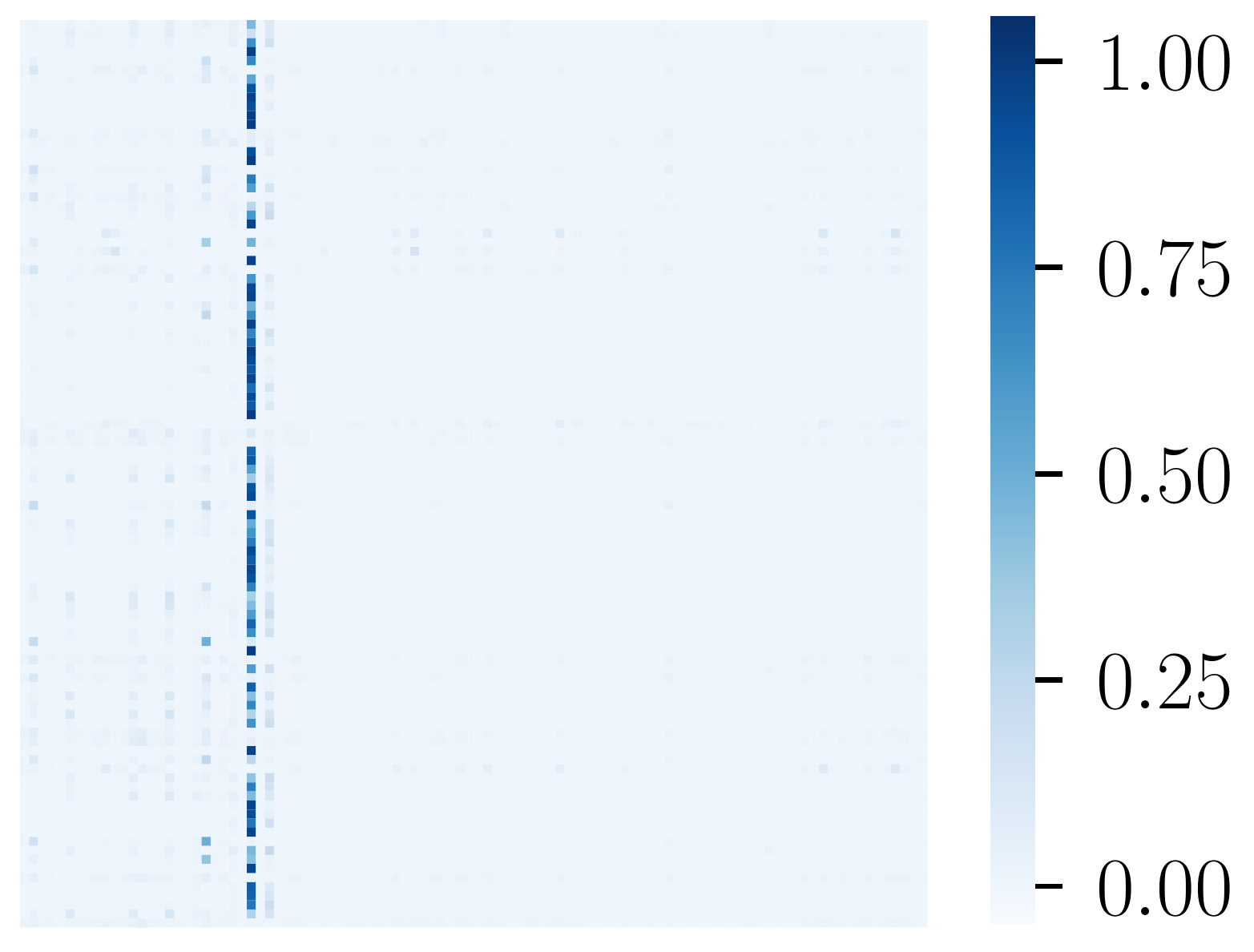}}    
    \subfigure[E-2, H-1]{
        \includegraphics[height=0.21\linewidth]{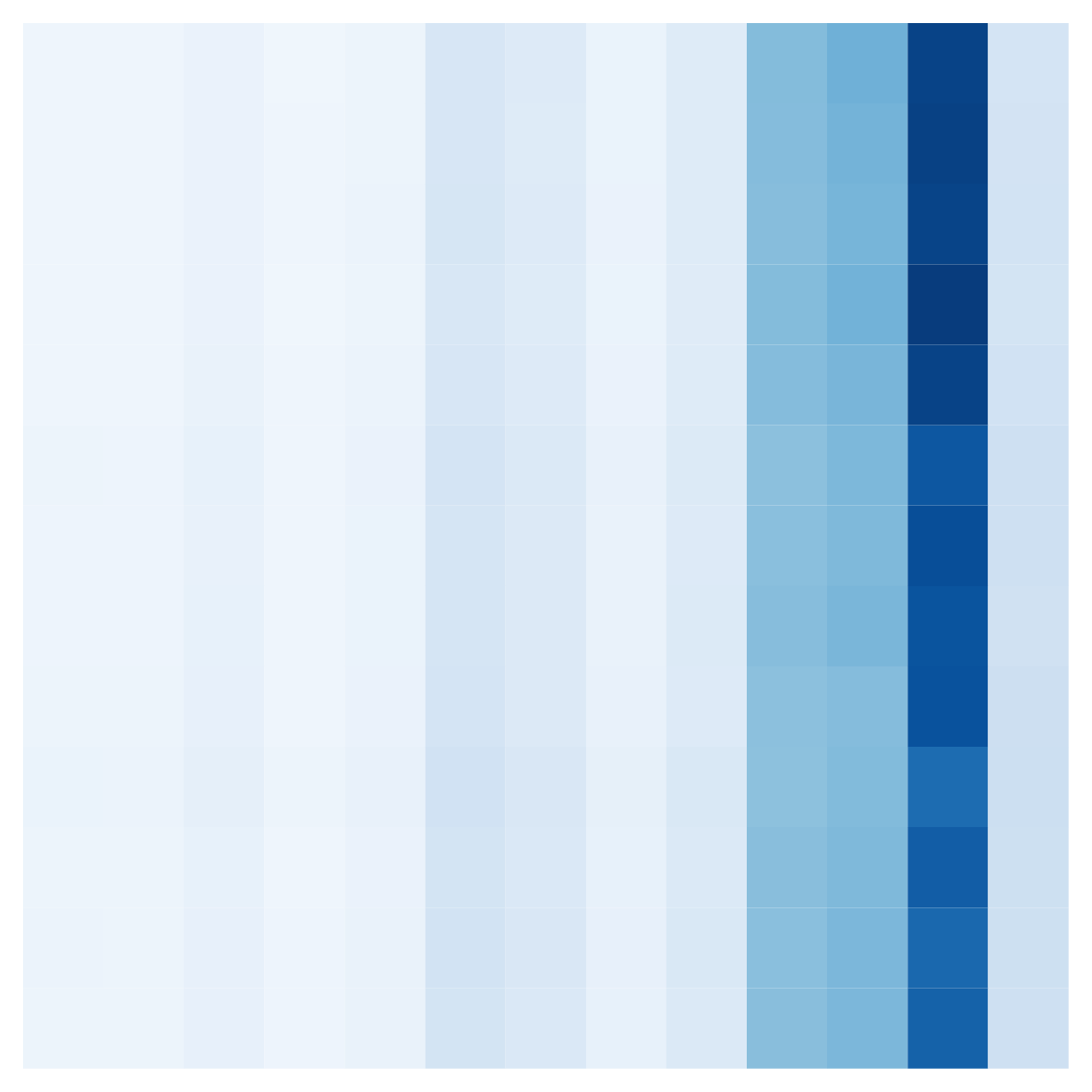}}
    \subfigure[E-2, H-2]{
        \includegraphics[height=0.21\linewidth]{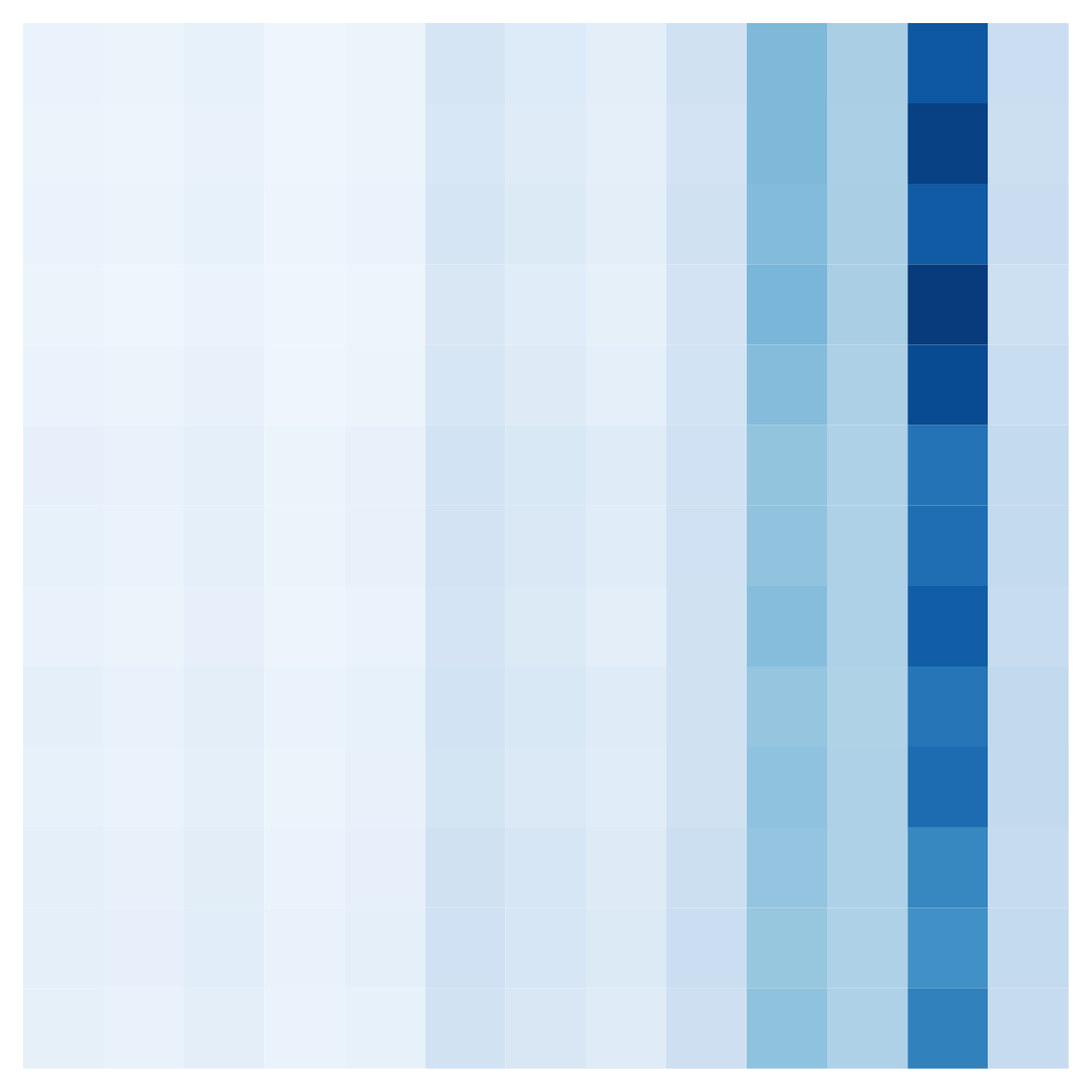}}
    \subfigure[E-2, H-3]{
        \includegraphics[height=0.21\linewidth]{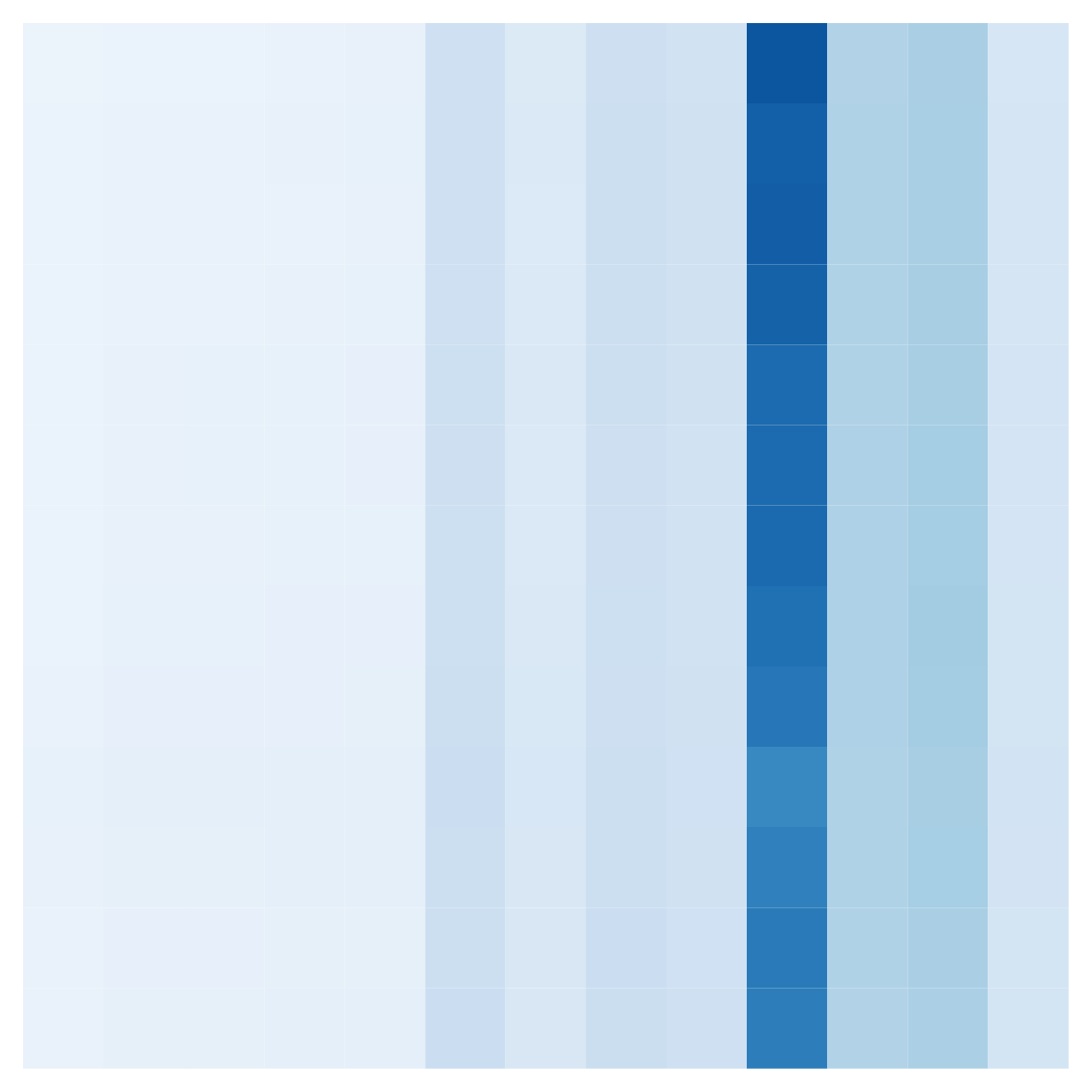}}
    \subfigure[E-2, H-4]{
        \includegraphics[height=0.21\linewidth]{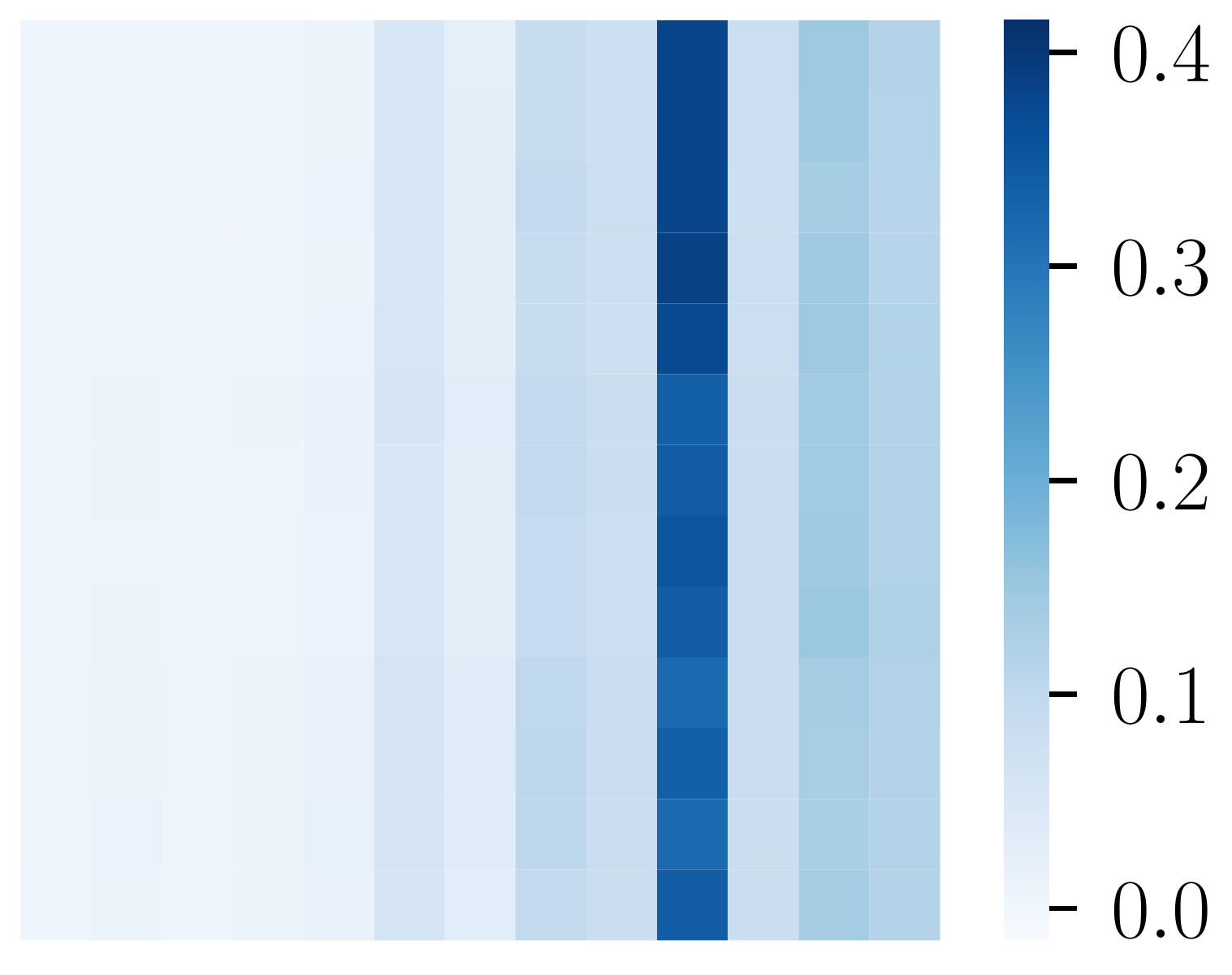}}    
    \caption{Heatmaps of global attention weights of a random sample (\textit{uid}: 172) from Foursquare NYC.}
    \label{fig:attentionmap}
\end{figure}

We first visualize the global attention weights of each head in each encoder to validate the capacity of STAR-HiT to encode spatio-temporal context. As shown in Figure \ref{fig:attentionmap}, different heads in Encoder-1 highlight different relevant check-ins, whereas heads in Encoder-2 after a round of subsequence aggregation consistently focus on the last few check-in subsequences, which seems to be the most representative subsequences correlated to the target check-in. Encoder-1 is responsible for encoding the correlations between individual check-ins, thus check-ins to the same POI or with similar spatio-temporal context to the target check-in are most likely to be assigned high weights. Note that even the weights of different check-ins to the same POI are not identical, due to the spatio-temporal context embedding that makes representations position-specific. Next, Encoder-2 focuses on subsequence-level correlation, thus the most relevant subsequences generally point to recent subsequences with similar periodicity.

\begin{figure*}[htb]
    \centering
    \begin{minipage}[t]{0.5\linewidth}
		\centering
		\subfigure[Correlation Matrix]{
		\includegraphics[width=0.99\linewidth]{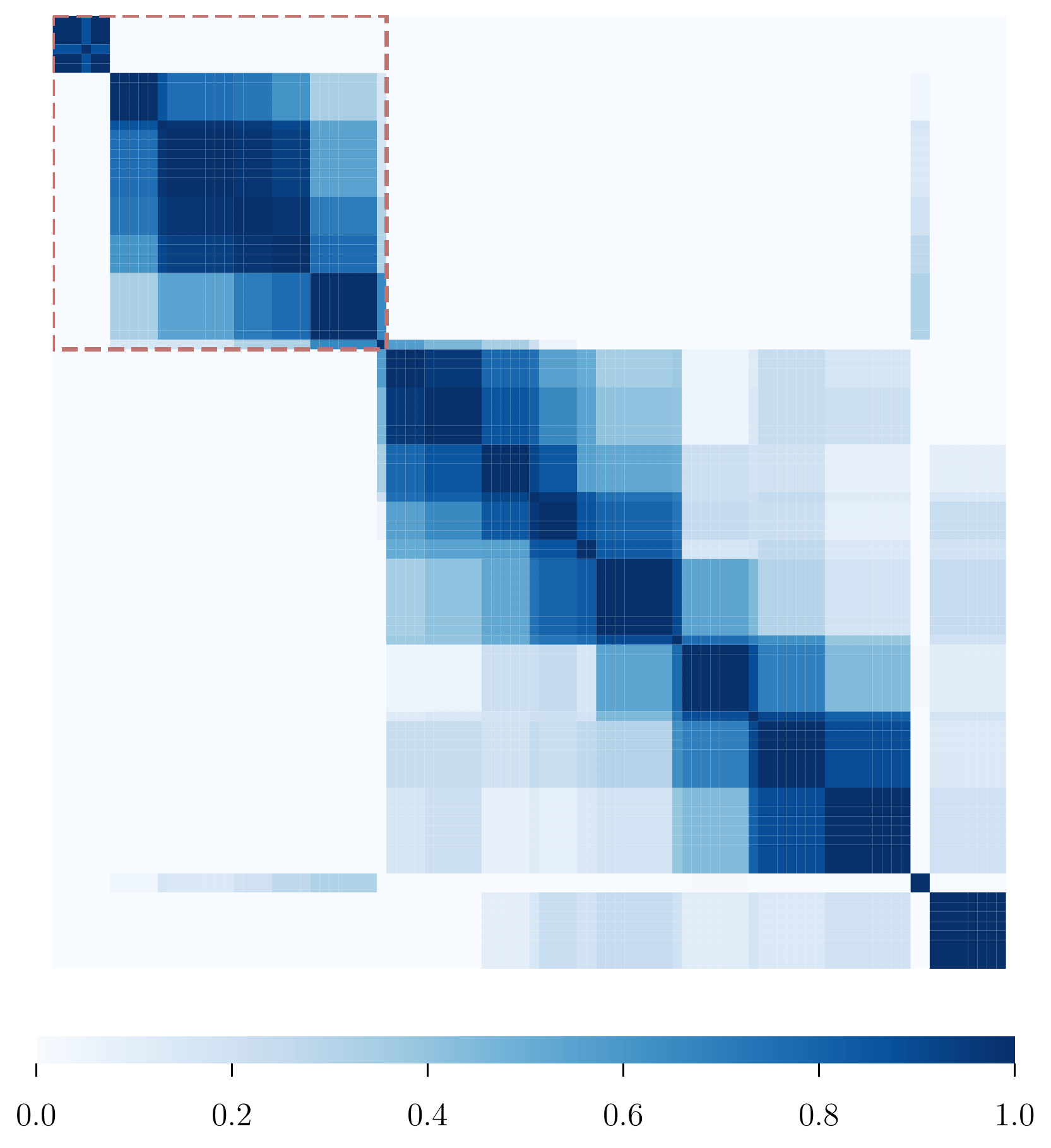}
		\label{fig:corr_mat}
		}\vspace{1pt}
		\subfigure[Check-in Sequence Segment]{
		\includegraphics[width=0.99\linewidth]{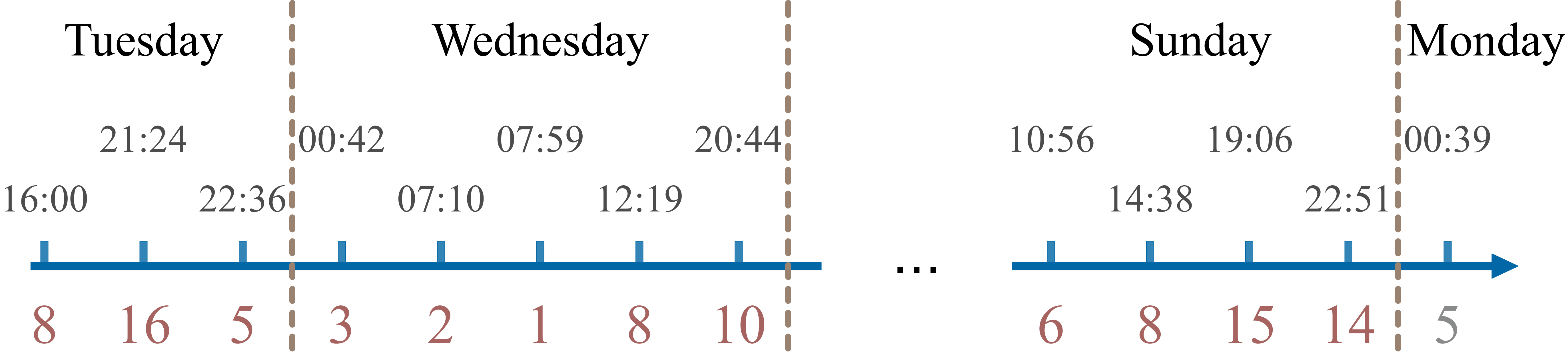}
		\label{fig:seq_seg}
		}
	\end{minipage}
	\begin{minipage}[t]{0.42\linewidth}
		\centering
		\subfigure[Map]{
		\includegraphics[width=0.907\linewidth]{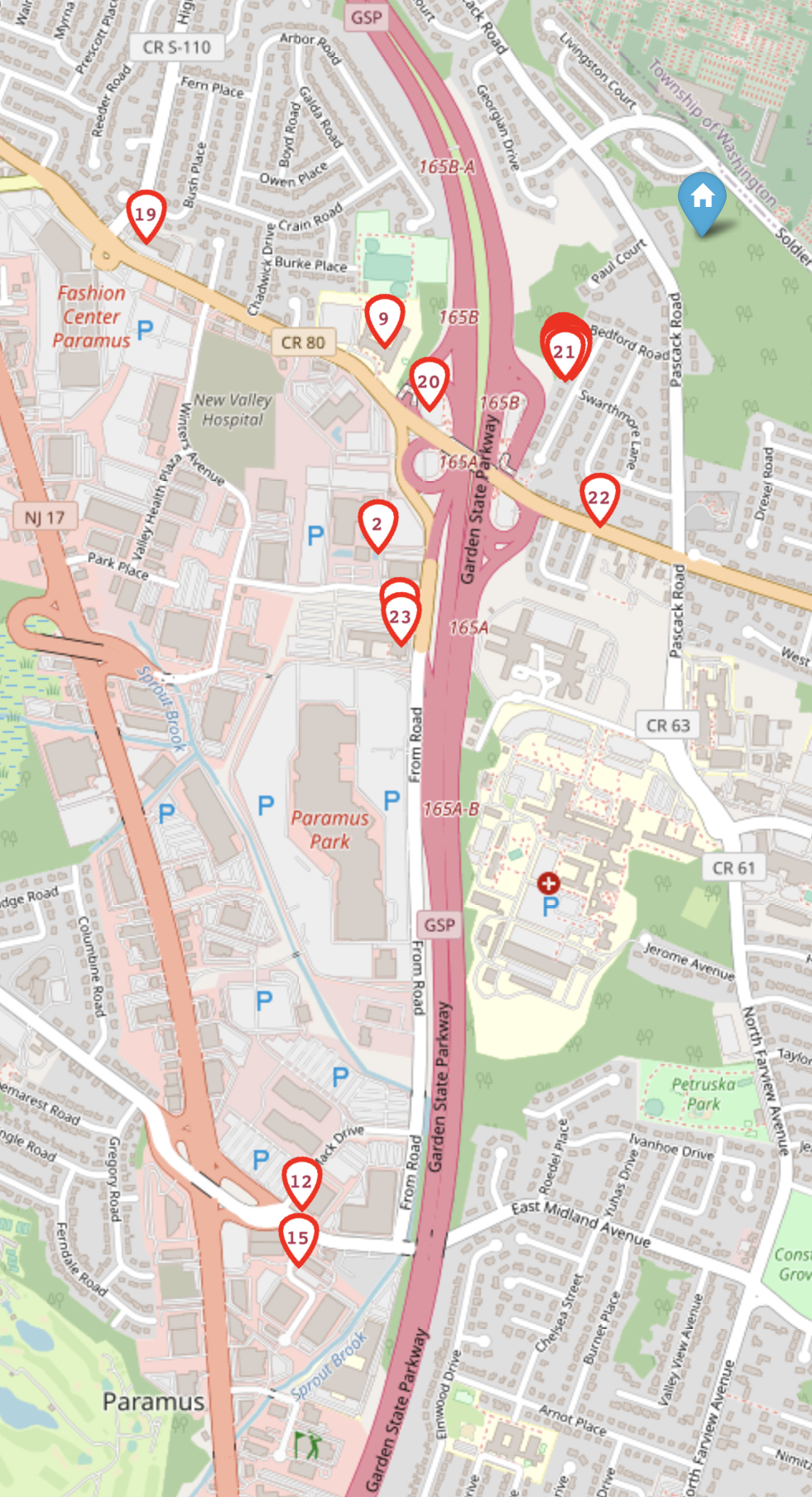}
		\label{fig:checkin_in_map}
		}
	\end{minipage}
\caption{Visualization of a random sample (\textit{uid}: 172) on the test set.}
\label{fig:vis}
\end{figure*}

In addition, we visualize the correlation matrix $M \in \mathbb{R}^{L\times L}$ of check-in representations to validate the effectiveness of STAR-HiT in capturing the latent hierarchical structure of the check-in sequence. The final check-in representations are obtained by: 1) concatenating the context-aware representations $\mathbf{E}(u)$ and their belonging subsequence representations in each encoder, 2) calculating the Pearson Correlation Coefficient (PCC) of normalized representations of $i$-th check-in and $j$-th check-in, which serve as each element $M_{i,j}$ in the correlation matrix. The correlation matrix is illustrated in Figure \ref{fig:corr_mat}, and meanwhile the check-in sequence segment corresponding to the region marked with the red-dotted rectangle is shown in \ref{fig:seq_seg}. We also plot the visited POIs projected on the map in Figure \ref{fig:checkin_in_map}. From Figure \ref{fig:vis}, we can see that STAR-HiT uncovers the multi-granularity semantic subsequences. For daily regularity, check-ins on Tuesday (\textit{pid}: 8, 16, 5) are identified as a subsequence, while the check-ins from Tuesday noon to Wednesday noon (\textit{pid}: 8, 16, 5, 3, 2, 1) are extracted as another subsequence. Furthermore, the weekly regularity is discovered as well, that is, the first week marked with the red-dotted rectangle in \ref{fig:corr_mat} is clearly separated from the following weeks. It should be noted that as shown in Figure \ref{fig:seq_seg}, some subsequences aggregate the last check-in (\textit{pid}: 5) with former check-ins while some do not. It is reasonable since that even though the check-in of POI 5 occurs on Monday, it occurs in the early hours of Monday, so it is more likely to be on the same itinerary as the previous check-in.

Overall, the proposed STAR-HiT can not only achieve superior recommendation performance, but also effectively capture the latent hierarchical structure present in check-in sequences, thereby providing explanations for personalized recommendations accordingly.

%% file: sections/5_conclusion.tex
\section{Conclusion}
In this work, we explore the latent hierarchical structure of sequential behavior patterns exhibited in user movements. We propose a novel Spatio-Temporal context AggRegated Hierarchical Transformer (STAR-HiT) model for next POI recommendation, which consists of stacked hierarchical encoders to capture the latent hierarchical structure of check-in sequences. Specifically, the hierarchical encoder is designed to jointly model spatio-temporal context and locate semantic subsequences with different positions and lengths in an explicit way. In each encoder, the global and local attention layers enhance spatio-temporal context modeling by capturing inter- and intra- subsequence dependencies; meanwhile, the sequence partition layer and subsequence aggregation layer adaptively locate and fuse semantic subsequences, and generate a new sequence with a higher level of granularity. This sequence is then fed into the next encoder for further subsequence discovery and sequence abstraction. By stacking multiple hierarchical encoders, semantic subsequences of different granularities are recursively identified and integrated, constituting the overall hierarchical structure present in the user movement. We perform extensive experiments on three public datasets to: 1) demonstrate that our proposed STAR-HiT outperforms state-of-the-art methods by a large margin, 2) get deep insights into the design of STAR-HiT, and 3) verify that STAR-HiT successfully captures multi-level semantic subsequences for revealing the latent hierarchical structure of the check-in sequence, so as to guarantee the robustness and explainability of the recommendation.